\documentclass[journal,draftcls,onecolumn,english]{IEEEtran}
\usepackage{amsmath,amssymb,amsfonts}
\usepackage{graphicx}
\usepackage{textcomp}
\usepackage{multirow}
\usepackage{caption}
\usepackage{subcaption}
\usepackage{hyperref} 
\usepackage{blindtext}
\usepackage{algorithm,algpseudocode}

\def\BibTeX{{\rm B\kern-.05em{\sc i\kern-.025em b}\kern-.08em
    T\kern-.1667em\lower.7ex\hbox{E}\kern-.125emX}}

\begin{document}

\title{Interplay between Cryptocurrency Transactions and Online Financial Forums}

\author{\IEEEauthorblockN{Ana Fernández Vilas}
\IEEEauthorblockA{\textit{atlanTTic Research Center, University of Vigo} \\
36310 Vigo, Pontevedra, Spain  avilas@det.uvigo.es \\}
\and
\IEEEauthorblockN{Rebeca P. Díaz Redondo}
\IEEEauthorblockA{\textit{atlanTTic Research Center, University of Vigo} \\
36310 Vigo, Pontevedra, Spain   \\}
\and
\IEEEauthorblockN{Daniel Couto Cancela}
\IEEEauthorblockA{\textit{atlanTTic Research Center, University of Vigo} \\
36310 Vigo, Pontevedra, Spain   \\}
\and
\IEEEauthorblockN{Alejandro Torrado Pazos}
\IEEEauthorblockA{\textit{atlanTTic Research Center, University of Vigo} \\
36310 Vigo, Pontevedra, Spain   \\}
}

\maketitle

%\and
%\IEEEauthorblockN{}
%IEEEauthorblockA{\textit{atlanTTic Research Center, University of Vigo} \\
%36310 Vigo, Pontevedra, Spain \\
%rebeca@det.uvigo.es}
%\and
%\IEEEauthorblockN{}
%\IEEEauthorblockA{\textit{atlanTTic Research Center, University of Vigo} \\
%36310 Vigo, Spain \\}
%\and
%\IEEEauthorblockN{Alejandro Torrado Pazos}
%IEEEauthorblockA{\textit{atlanTTic Research Center, University of Vigo} \\
%36310 Vigo, Spain \\}
%}

\maketitle

\abstract{ CCryptocurrencies are a type of digital money meant to provide security and anonymity while using cryptography techniques.  Although cryptocurrencies represent a breakthrough and provide some important benefits, their usage poses some risks that are a result of the lack of supervising institutions and transparency. Because disinformation and volatility is discouraging for personal investors, cryptocurrencies emerged hand-in-hand with the proliferation of online users' communities and forums as places to share information that can alleviate users’ mistrust. This research focuses on the study of the interplay  between these cryptocurrency forums and fluctuations in cryptocurrency values. In particular, the most popular cryptocurrency Bitcoin (BTC) and a related active discussion community, Bitcointalk, are analyzed. This study shows that the activity of  Bitcointalk forum keeps a direct relationship with the trend in the values of BTC, therefore analysis of this interaction would be a perfect base to support personal investments in a non-regulated market and, to confirm whether cryptocurrency forums show evidences to detect abnormal behaviors in BTC values as well as to predict or estimate these values. The experiment highlights that forum data can explain specific events in the financial field. It also underlines the relevance of quotes (regular mechanism to response a post) at periods: (1) when there is a high concentration of posts around certain topics; (2)~when peaks in the BTC price are observed; and, (3) when the BTC price gradually shifts downwards and users intend to sell.}


%%%%%%%%%%%%%%%%%%%%%%%%%%%%%%%%%%%%%%%%%%

%%%%%%%%%%%%%%%%%%%%%%%%%%%%%%%%%%%%%%%%%%

\section{Introduction}
\label{sec:introduction}

Cryptocurrencies~\cite{StateApproachesCryptocurrenciesarticle,BacktoBasics} are a type of digital money meant to provide security and anonymity while using cryptography techniques, so that transactions are not certified or verified by intermediaries such as central banks but  directly by users throughout a decentralized peer-to-peer systems. Although cryptocurrencies represent a technology breakthrough and bring some important benefits, the aforementioned absence of supervising institutions and regulations to protect users poses some risks~\cite{FATF-OECD,AML/CFTRisks,WorldBank} that cannot be ignored.

On the one hand, disinformation can mislead users in the assessment of the investment benefits when buying cryptocurrencies and, on the other hand, the emergence of fraudulent activities in cryptocurrency markets can intentionally cause  losses.  As a consequence, price volatility, i.e., big fluctuations in  exchange rates, is a common scenario that directly affects cryptocurrency holders when exchanging the virtual currency into a real one. Altogether, these risks turn cryptocurrencies' investment into a bet in terms of profit. Therefore, cryptocurrencies may be discouraging for personal investors who have no or little knowledge about them~\cite{european2015virtual,BankInternationalSettlements}, and they may be seen as a speculation vehicle instead of a payment form.  Given apart the risks that are related to assess investment benefits: disinformation, fraudulent activities, and volatility; cryptocurrencies' legal status also produces confusion regarding rights and  responsibilities. Users must be aware  that there is no chance of a refund for  a product or service, even if the received product or service is not what was expected. Besides, in the case of a faulty transaction (e.g., mistakes on amounts or beneficiaries' identities), it cannot be undone because of users' anonymity and the absence of a recognized authority~\cite{2VirtualCurrency}. 

With all the above, the cryptocurrency market can be described as one where users have no confidence in each other or in the currency they are using. For this reason, cryptocurrencies emerged hand-in-hand with the proliferation of user communities and forums as places to share  information that can alleviate users' mistrust. In the hostile cryptocurrency market, online financial forums would be valuable and informative, provided that investors could rely on their credibility. Although financial forums credibility can be approached in a similar way to other social networks, the main challenge of this paper is obtaining a first sign of the informative power of cryptocurrency forums by  checking whether financial forum activity is aligned with cryptocurrency dynamics over time.

This work focuses on Bitcoin (BTC), the most widely known cryptocurrency on the market, in order to solve this research question. We deployed an experiment to analyze the activity in cryptocurrency forums, Bitcointalk in particular, and its impact on fluctuations and trends of BTC price. Bitcointalk is a big and complex community of cryptocurrency users: a hierarchy with a fixed number of ranks and a rewarding mechanism based on merit and activity points to level up in the hierarchy. From the data that were gathered from Bitcointalk and the BTC values over the study period, the contribution of this paper focuses on analyzing the impact of forum activity on BTC values. For this, both Bitcointalk activity and BTC prices are represented as time series to apply point-to-point and windowed analysis. This contribution is located in the framework of our  previous works~\cite{DBLP:journals/eswa/EvansOCV19, IntelliSys2018, DBLP:journals/mta/VilasRCOE19}, where a data fusion model is defined that gathers not only social media, but also other data sources, like stock market data, regulatory new services, broker agencies, etc. The ecosystem will provide a dashboard for small investors, who, unfortunately, do not have access to high quality information and can be misled by contributions in social media. This work pursues the inclusion of social media as evidence for setting up an ecosystem to detect potential irregularities in the cryptocurrency market.  Specifically, this research focuses on the information collected from Bitcointalk as cryptocurrency forum.

This paper is organized, as follows. Section~\ref{sec:related} reviews related work in the application of behavioral sciences to analyze cryptocurrency dynamics; and, Section~\ref{sec:forums} describes the most popular cryptocurrency forums. After the selection of a cryptocurrency forum with good characteristics to obtain a relevant dataset of users' BTC comments, a graphical overview of the cryptocurrency landscape and the main steps in the experiment are shown in \mbox{Section~\ref{sec:overview}}: the extraction of the BTC value for the period of study (Section~\ref{sec:data:extraction}); exploratory analysis of the characteristics of the forum in order to select the features that are more predictive in terms of the BTC value (Section~\ref{sec:exploratory:analysis}); study of the relationship between Bitcointalk activity and fluctuations of BTC price (Section~\ref{sec:results}); correlation and predictive power of Bitcointalk \mbox{(Section~\ref{sec:predictive:power}).} The results  shed light on the study matter, the informative potential of financial forums in a deregulated and misinformed market. Finally, Section~\ref{sec:discussion} includes extra comparison with other cryptocurrencies, as well as some issues that are related to the period of study, which deserves further research; and, Section~\ref{sec:conclusions} summarizes conclusions and future work.

\section{Related Work}
\label{sec:related}

%Blockchain

Because cryptocurrencies do not impose users' financial statements or identification requirements, reliability fully remains on the security of the  peer-to-peer system and the Blockchains technology ~\cite{10.1145/3366370}: data structures that are packed in ledgers that store and exchange data in each pack or block.  Whenever a node creates a new block,  an encrypted broadcast message is sent to every single node in the network, so that the information only remains available to the peers. The receivers of this broadcast message start security processes to validate and confirm the validation of the new block. Afterwards, every single node adds the block to its ledgers and communicates the change to every group in the network. This results in a distributed network, where all of the members create the block chain by storing the same data with high quality in terms of transparency and security.  On the other hand, the mechanism to reach a consensus on which nodes will be included in the chain depends on different kinds of proof: proof of work (PoW), proof of stake (PoS), and proof of importance (PoI). Although Blockchain technology produces a structure of data with inherent security qualities to be applied to a variety of sectors~\cite{CASINO201955} (financial, health, education, etc.), cryptocurrency markets continue being out of supervision by governments or a world-wide economic institution, which poses risks for cryptocurrency users and community.

These new risks in the financial sector area has received a lot of attention in the data analysis community, with the aim of reproducing previous achievements in regulated markets. A plethora of academic works study the capability of social channels to predict returns and detect abnormal behaviors on the stock market, \cite{doi:10.1111/deci.12229, DERAKHSHAN2019569}, to cite a few. Regarding the predictive power of cryptocurrencies, in~\cite{CATANIA2019485},  authors report promising results in the prediction of the most capitalized cryptocurrencies (Bitcoin, Litecoin, Ripple, and Ethereum) based on predictors that are external to cryptocurrency trading, specifically, commodity prices, stock and bond prices, and volatility indices. An alternate approach proposed in~\cite{Bohte-Rossini-2019} also studies the forecasting ability of cryptocurrencies with the idea of crypto-predictors in~\cite{CATANIA2019485}, but considering time-varying volatility in the cryptocurrency time series. More recently,~\cite{Bianchi162} also focuses on the relationship of cryptocurrency returns and volatility with traditional assets and provides insights that suggests that, as in traditional markets, investors’ sentiment play an important role in trading. These works confirm a trend in the behavioral alignment of cryptocurrencies with traditional markets. Accordingly, if social media has been acknowledged as an important factor in regulated markets, its role in cryptocurrency market should be~studied. 

 In this research area, our previous work~\cite{DBLP:journals/mta/VilasRCOE19} applies anomaly detection to  validate the relevance of Twitter as a sign for stock-impacting financial events, despite the ambiguity issue in Twitter hashtags linked to companies in stock markets, as we analyzed in~\cite{DBLP:journals/eswa/EvansOCV19}. More recently, cryptocurrencies have become a major phenomenon, both the cryptocurrency itself as a commodity and the technology that cryptocurrency development has brought~\cite{MALLQUI2019596, 10.3389/fbloc.2020.00001}. 

Nevertheless, as there are no regulations to protect users,  cryptocurrencies pose some risks. The lack of transparency and supervising institutions can lead to fraudulent activities, since it is not clear who gives information to users and how, so that fraudulent transactions, fake information, etc. can mislead users in analyzing their risk and, as a consequence, can produce losses. Additionally, the uncertainty of their legal status creates confusion regarding rights and legal responsibilities, and the big fluctuations in exchange rates directly affect the holders when exchanging the virtual currency into real one. Therefore, cryptocurrencies are seen by many as a speculation vehicle instead of a form of payment. Taking all of this into account, the cryptocurrency market can be described as one where users have no confidence in each other or in the currency they are using. For this reason, since their creation, lots of user communities and forums that are related to transactions and currencies emerged as a place to meet and share information. 

Given that investors utilize social media, i.e., financial forums, as a means of sharing news about international stock exchanges and, more recently, cryptocurrencies, behavioral sciences and related scientific literature have researched evidences of the relationship between social media and price fluctuations of cryptocurrencies, and even of the kind of products cryptocurrencies that are planned to be spent on~\cite{RePEc:prg:jnlaip:v:2019:y:2019:i:2:id:127:p:112-138}. Several cryptocurrency forums, such as ADVFN, Moonforum, Blackhatworld, Bitcointalk, Cryptocompare, etc. have been used in this kind of research. 

The literature~\cite{steinert2018predicting} suggests a relationship between social media and prices for both cryptocurrencies and stocks. Given the extreme volatility in cryptocurrency prices,~\cite{10.3389/fbloc.2020.00001} proposes a framework to discover  potential causes of shifts in the price movement that are captured by social media discussions by grouping words of similar meaning to identify the underlying concepts.  With Reddit as data source,  the  framework identified plausible evidences for the shifts in bitcoin price trends, as well as in ether. BTC is also the focus in~\cite{RePEc:prg:jnlaip:v:2019:y:2019:i:2:id:127:p:112-138}, which studies goods or services purchased using  Bitcointalk forum as evidence.  Besides the main conclusion, i.e.,  cryptocurrencies are used to buy and sell at the electronics and computer segments, a script was designed for  Bitcointalk, as in our work, to go through  advertisements and download relevant data.  Additionally, in~\cite{steinert2018predicting}, the relationship between social media and cryptocurrencies' price fluctuations was studied for smaller currencies, i.e., cryptocurrencies with small trading volumes so-called altcoins. Based on their  findings,  altcoin returns can be predicted  by using the information provided by Twitter as datasource.

\section{Cryptocurrency Forums}
\label{sec:forums}

The first step is to select one of them as information source in order to analyze the activity (comments, posts) that the users publish online in cryptocurrency forums. The exploration of the deep web---the part of the web invisible to standard search engines---should be taken into consideration due to the anonymity inherent to cryptocurrencies. However, content in the Deep Web is still discarded, since the study focuses on open forums, which are accessible to everyone willing to participate, and avoiding those linked to illicit or illegal activities. In order to select a cryptocurrency forum from the surface web, we consider the potential volume of data, but, as importantly, that the information can be retrieved. Regarding the latter,  the forum site should offer a specific API to data retrieval; otherwise, the information should be gathered through crawling frameworks.  In the following paragraphs, we describe the most popular  forums, precisely,  ADVFN , Moonforum , Blackhatworld , Bitcointalk,  and Cryptocompare,  to find a data source that can be automatically mined and, at the same time, relevant enough to  BTC and our research questions. 

ADVFN~(\url{https://uk.advfn.com}) is dedicated to the financial market and it offers a wide amount of information in real time regarding cryptocurrencies: actual value, maximum and minimum rates, evolution, etc. Among its multiple options and sections, it has its own forum. This does not have any kind of organization or hierarchy and the threads deal with very diverse issues, such as the effects of politics on economics or the evolution of the oil market. At the moment of the analysis, a search by the names of the most famous cryptocurrencies enabled a classification of the threads into topics: 15 topics were about BTC and the total volume was around 7000 posts; two topics were about Ethereum, with 31 posts in total; two topics were about Ripple, with 37 posts in total. Filtering cryptocurrency threads is not a simple task. Using related vocabulary, like “mining”, in the search, hundreds of results are returned, although most of them are about mining materials, such as silver or gold. This adds the difficulty of identifying what topics are really about cryptocurrencies. In spite of that, it is convenient to remark that users include graphics and hyperlinks in their posts, which proves the rigor of the forum. With respect to data extraction, ADVFN offers many APIs to give access to its financial data, but it does not have one to access posts. Still, data extraction is possible by crawling the website.

{Moonforum}~(\url{https://www.moonforum.net)} is a cryptocurrency forum that counted with 40,447 posts when analyzed. It has an organization structure with thee levels, going from general to specific: categories, boards, and topics. This forum is characterized by having lots of boards where users make subjective assessments and share their experiences. However, it is a small forum: some of the boards only have a few topics or are completely empty. It does not have an API to access data, so the information should be collected with crawling techniques. 

Blackhatworld~(\url{https://www.blackhatworld.com}) is a forum that is focused on online marketing where users discuss communication techniques in social media, graphic design, and the development of web applications. It has a section that is dedicated to cryptocurrencies with approximately the same volume of posts as Moonforum. There is no organizational structure in it, and users share their opinions and seek advice from each other. Data extraction is only possible by website crawling. 

Bitcointalk~(\url{ https://bitcointalk.org}), according to their own statistics, has 2,460,321 users, 47,227,046 comments from 2009 to the date of the analysis and an average of 8033 new posts every day. Its organizational structure is similar to Moonforum. It is focused on bitcoin, but includes a category for the rest of cryptocurrencies. In the forum, users share their opinions and experiences, with an important degree of participation and potential information. There is an unofficial API to extract the posts; but, retrieving the information by web crawling is also possible.

Cryptocompare~(\url{https://www.cryptocompare.com}) is a website that provides information and analysis about cryptocurrencies and has a forum section for all of them. It has multiple APIs to access the analysis data they offer, but none of them gives access to posts. Its design that is based on Angular framework makes Website crawling a hard task.

After considering all forums,  Bitcointalk was selected, because it offers higher levels of activity, which guarantees a convenient  volume for the analysis; and, it is exclusively devoted to cryptocurrencies, which guarantees the relevance of the data for our study.

 \section{Methodology}
 \label{sec:overview}

\begin{figure}[H] %\centering
\includegraphics[width=0.8\columnwidth]{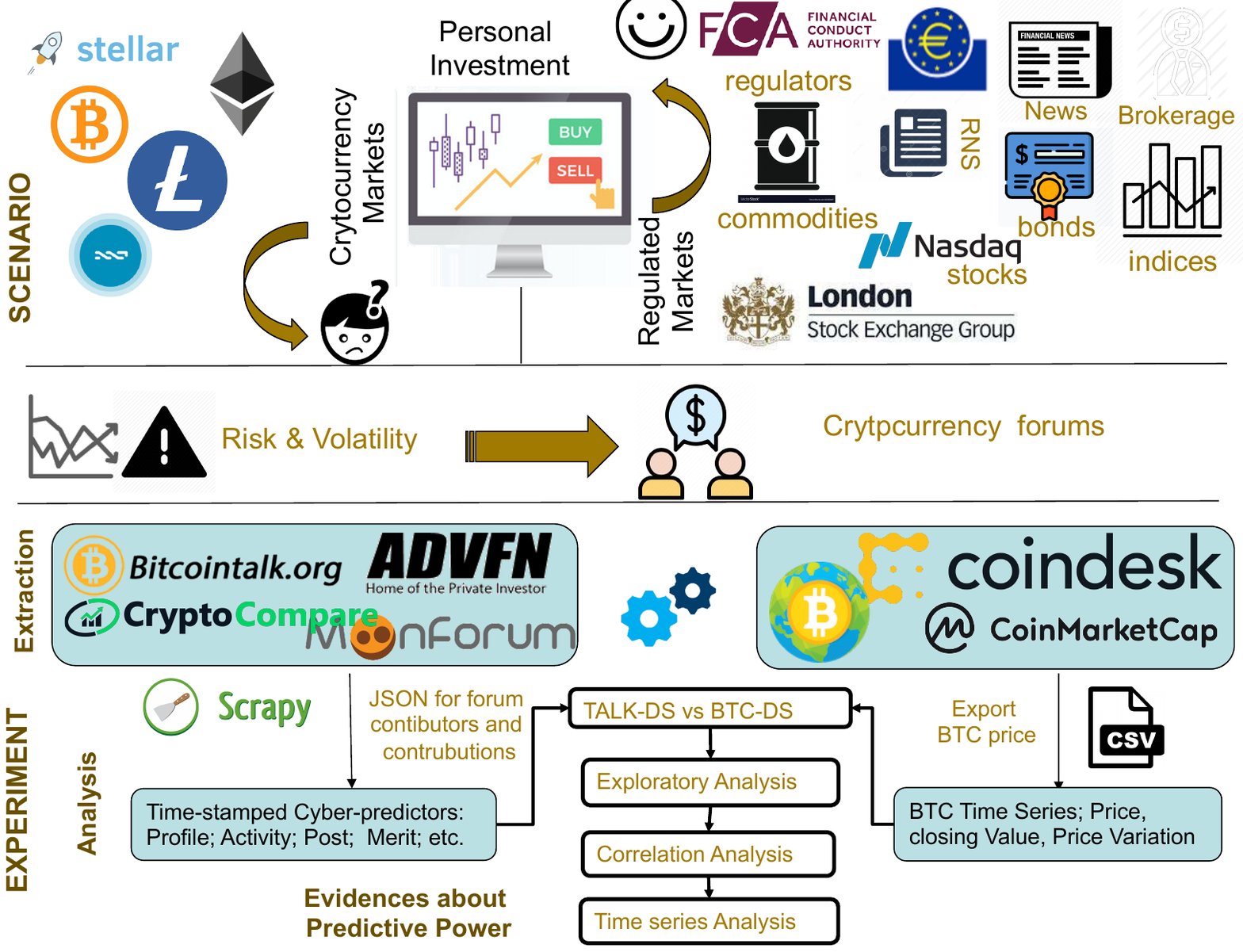} \caption{Overview of the experiment.} \label{fig:overview} \end{figure}

Figure~\ref{fig:overview} provides a general overview of the scenario, its challenges, and the main objectives in this research work. A regulated/controlled market is one in which the government or another public authority exercises some degree of oversight and facilitates multiple third-party buying and selling interests in financial instruments. Commodities, bonds, and stocks are traditional considered regulated markets in accordance with their non-discretionary rules. On the contrary, cryptocurrencies function without institutional backing and they are intrinsically borderless and, as a consequence, in a high volatile scenario which may impede their usage for payments or settlement.  Accordingly, investors, particularly personal investors, may not be prepared to rapid boom/bust cycles. Unlike regulated markets, crypto-asset trading platforms do not have measures to mitigate price swings. Higher risk can be alleviated by different data sources to allow a personal investor to take an informed decision. In this respect, cryptocurrency investors simply rely on online forums and the own cryptocurrency price analysis as information sources.

The experiment in this paper studies cryptocurrency forums to uncover the dynamics of their activity and their relationship with a fluctuation in cryptocurrency prices in order to check whether cryptocurrency forums can be a support tool to face risk and volatility. Although we cannot demonstrate causality without a controlled experiment, a response to this research question can give us some guarantees regarding the credibility of the activity in these forums and their usage as a sound information source for individual and amateur investors.  It should be remarked that, even if there is no government or financial entities that regulate cryptocurrencies, users are required to pay taxes for all BTC processes that have a value, and they are responsible for searching and assuming all of the legal procedures to pay these taxes. Not only does volatility make investment a bet, but users may also commit a financial offense.

This work focuses on Bitcoin, the most widely known cryptocurrency in the market, in order to solve this research question. We deployed an experiment to analyze the activity in cryptocurrency forums and its correlation with the fluctuation of BTC price. The research question that is required to retrieve information from the Bitcointalk as well as BTC values, so that we can obtain two datasets: TALK-DS for Bitcointalk activity and BTC-DS for BTC values. For Forum-DS, Bitcointalk API was discarded in favor of scrapping techniques, because it is an unofficial API, not maintained by the platform, so that vulnerabilities can arise in terms of errors, incoherent results by possible flaws in its logic or even pose a threat to security and privacy. Accordingly, the extraction was carried out with the framework Scrappy by developing three spiders (classes to parse and extract data from websites): one profiles’ spider; posts’ spider; and topic’s spider, which keep the intended data in Python dictionaries.  The methodology consider social media predictors for cryptocurrencies, which is, a set of cryptocurrency forums time stamped features that are analyzed along the BTC time series, according to Figure~\ref{fig:overview}. The study was carried out from simplest exploratory analysis to point-to-point correlation analysis and, finally, time series analysis based on cross-correlation.

This paper is our most recent contribution in the field of the role of financial forums in regulated and not regulated markets. Regarding regulated markets we have published different results that are related to the London Stock Exchange (LSE)~\cite{DBLP:journals/eswa/EvansOCV19, IntelliSys2018, DBLP:journals/mta/VilasRCOE19}. These works pursue the definition of an ecosystem that is based on data fusion (data sources: Live market data, RNS, Twitter, Broker Agencies, company web sites, etc.). Because all of the ecosystem is located in a regulated market (LSE), the work in this new paper also pursues the inclusion in our ecosystem of non-regulated markets. In this respect, we have researched on the relevance of Twitter as a financial source of information for cryptocurrency trading ~\cite{DBLP:journals/access/RG20}.

\section{Dataset}
\label{sec:data:extraction}

The retrieved information gathers common data on user profiles and posts (usernames, number of posts, etc.), but also some other specific features that are related to Bitcointalk (marked with * in Table~\ref{tab:information:bitcointalk}).

\begin{table}[H]

\caption{Retrieved Information from Bitcointalk.}
\label{tab:information:bitcointalk}
\begin{tabular}{llllll}
%\tablesize{\scriptsize}
%\setlength{\cellWidtha}{\columnwidth/6-2\tabcolsep+0.250in}
%\setlength{\cellWidthb}{\columnwidth/6-2\tabcolsep+0.0in}
%\setlength{\cellWidthc}{\columnwidth/6-2\tabcolsep-0.0in}
%\setlength{\cellWidthd}{\columnwidth/6-2\tabcolsep-0.0in}
%\setlength{\cellWidthe}{\columnwidth/6-2\tabcolsep-0.15in}
%\setlength{\cellWidthf}{\columnwidth/6-2\tabcolsep-0.1in}
%\scalebox{1}[1]{\begin{tabularx}{\columnwidth}{>{\PreserveBackslash\raggedright}p{\cellWidtha}
%>{\PreserveBackslash\raggedright}p{\cellWidthb}
%>{\PreserveBackslash\raggedright}p{\cellWidthc}
%>{\PreserveBackslash\raggedright}p{\cellWidthd}
%>{\PreserveBackslash\raggedright}m{\cellWidthe}
%>{\PreserveBackslash\raggedright}m{\cellWidthf}}
%\toprule
\multicolumn{4}{l}{\textbf{User Profile}} & \multicolumn{2}{l}{\textbf{Posts}} \\ 
Username	&Number of Posts&	Last active & Gender & Username	 & Date and time  \\ 
Signature	& Bitcoin address &	 Age & Status & Topic Name & Emojis 	 \\ 
 Information contact	 & Topic Name & Location	& Date registered & Post \\ 
\multicolumn{6}{l}{\textbf{Bitcointalk specific}}  \\ 
Activity~*	& Position~*	& Merit~*	& Posts List~* & Quotes~* 	&\\ 
	Started Topics List~* &  &&&& \\

\end{tabular}
\end{table}

In the following paragraphs, we give a proper description of the extracted features. (1)~Activity measures the  participation in Bitcointalk, being regulated by forum administrators, which is automatically updated to all users every two weeks. The maximum number of points be earned in one period is 14, which implies an average of one post per day. (2)~Merits are rewards that users give to each other when one of them makes a valuable contribution to the forum. They are similar to social “likes” but with some restrictions, which is, only administrators and special users can give merit points without restrictions; for regular users to give one merit points they need to have earned two of them; (3)~Position classifies users according to their activity and merit, so that higher positions have more privileges (signatures, avatar, etc.) and less restrictions in term of posting limits, private messages, etc. Administrators, Moderators, Global Moderatos, and Staff are special user profiles without restrictions; also paid memberships are allowed to skip restrictions; (4)~Posts List contains all posts published by one user in the entire forum; (5)~Started Topic List contains posts that were published by a user that open a topic; and, (6)~Quotes contains chains of messages inside the body of one post created when users quote other members to keep track of a specific conversation. Finally, users are ranked in a Position according to their degree of participation in the forum. Table~\ref{tab:ranks:bitcointalk} describes the requirements that are needed to be assigned a specific position.

\begin{table}[H]
%\tablesize{\small}
\caption{Ranks in  Bitcointalk. \label{tab:ranks:bitcointalk} }
%\setlength{\cellWidtha}{\columnwidth/3-2\tabcolsep+0.0in}
%\setlength{\cellWidthb}{\columnwidth/3-2\tabcolsep+0.0in}
%\setlength{\cellWidthc}{\columnwidth/3-2\tabcolsep+0.0in}
%\scalebox{1}[1]{\begin{tabularx}{\columnwidth}
%{>{\PreserveBackslash\centering}m{\cellWidtha}
%>{\PreserveBackslash\centering}m{\cellWidthb}
%>{\PreserveBackslash\centering}m{\cellWidthc}}
%\toprule
\begin{tabular}{lll}
\textbf{Rank}	&\textbf{Activity}	&\textbf{Merit}	 \\ 
Brand New	&0	&0	\\
Newbie	&1	&0	\\
Jr. Member	&3	&1	\\
Member	&6	&10 \\
Full Member	&120	&100 \\
Sr Member	&240	&250\\
Hero Member	&480	&500 \\
Legendary&	Random, more than 775	& 1000\\

\end{tabular}
\end{table}

The viability of the extraction process was checked according to the number of Bitcointalk HTTP requests, which implies a significantly high execution time for the retrieving process and the risk of identifying the extraction process as a Denial of Service attack (suspicious high-density traffic from the same specific source). The extraction process was designed according to the maximum frequency  Bitcointalk HTTP requests allowed by introducing inactive periods. An inactive period of 0.6 seconds was used to not overload the website. At the time of data extraction, there would be 2.4 million pages of posts and 600,000 pages of topics in round numbers, so that a 0.6 inactive time for each profile and page would have resulted in 38,194 days in waiting time. A sample dataset was extracted, as retrieving all the data would take several months with a single machine and also to  not disturb Bitcointalk operation. When considering  the size of the community (2.5 million user profiles), the sample  population for TALK-DS was fixed to 1\% of the active community; a profile is considered to be active if it has at least one post during the study period. 

On the other hand, for BTC-DS, the cryptocurrency values should also be obtained to analyze whether there is any correlation between social media activity and trade activity. As cryptocurrencies, BTC in particular, there are plenty of widely known APIs to retrieve accurate currency values in real time, e.g., CoinMarketCap, BuyBitcoinWorldwide, and Coindesk . The values for this study were extracted by Coindesk because of the convenience of its output. For every day in the year of study, we obtained the opening prices, the closing price, and lowest and highest prices during the day (example in Table~\ref{tab:coindesk}).

\begin{table}[H]
%\tablesize{\scriptsize}
\caption{Extraction model of Bitcoin statistics in Coindesk. \label{tab:coindesk} }
%%\setlength{\cellWidthb}{\columnwidth/6-2\tabcolsep+0.0in}
%\setlength{\cellWidthc}{\columnwidth/6-2\tabcolsep-0.0in}
%\setlength{\cellWidthd}{\columnwidth/6-2\tabcolsep-0.0in}
%\setlength{\cellWidthe}{\columnwidth/6-2\tabcolsep-0.0in}
%\setlength{\cellWidthf}{\columnwidth/6-2\tabcolsep-0.0in}
%\scalebox{1}[1]{\begin{tabularx}{\columnwidth}{>{\PreserveBackslash\raggedright}p{\cellWidtha}
%>{\PreserveBackslash\raggedright}p{\cellWidthb}
%>{\PreserveBackslash\raggedright}p{\cellWidthc}
%>{\PreserveBackslash\raggedright}p{\cellWidthd}
%>{\PreserveBackslash\raggedright}m{\cellWidthe}
%>{\PreserveBackslash\raggedright}m{\cellWidthf}}
%\toprule
\begin{tabular}{llllll}
\textbf{Currency}	& \textbf{Date}		& \textbf{Closing Price} &	\textbf{24 h Open} &	\textbf{24 h High}&	\textbf{24 h Low} \\ 
BTC	01/01/2018& 	T05:00:00.000Z	& 13439.4175 &	13062.145 &	14213.44125	& 12587.60375\\
BTC	02/01/2018&	T05:00:00.000Z&	13337.62125&	13439.4175&	13892.2425&	12859.8025\\
BTC	03/01/2018&	T05:00:00.000Z&	14881.545	&13337.62125&	15216.75625&	12955.965\\
BTC	04/01/2018&	T05:00:00.000Z&	15104.45	&14881.545&	15394.98625&	14588.595\\
BTC	05/01/2018&	T05:00:00.000Z&	14953.8525	&15104.45&	15194.40625	&14225.16625\\
BTC	06/01/2018&	T05:00:00.000Z&	16576.69625	&14953.8525&	17118.35625&	14816.50875\\
BTC	07/01/2018&	T05:00:00.000Z&	16735.10625	&16576.69625&	17211.92375&	16254.94875\\
BTC	08/01/2018&	T05:00:00.000Z&	15632.41125	&16735.10625&	16861.20875&	15546.415\\
BTC	09/01/2018&	T05:00:00.000Z&	15242.93625	&15632.41125&	15944.5225&	13957.9125\\
BTC	10/01/2018&	T05:00:00.000Z&	14122.825	&15242.93625&	15360.1325&	13968.24375\\
BTC	11/01/2018&	T05:00:00.000Z&	13269.5975	&14122.825&	14942.61125&	13161.5775\\
BTC	12/01/2018&	T05:00:00.000Z&	13633.36375	&13269.5975&	14198.5325&	12845.7125\\ 

\end{tabular}
\end{table}

\section{Exploratory Analysis}
\label{sec:exploratory:analysis}

TALK-DS consists of 24,033 profiles (active during 2018, although the total number of profiles inspected was higher, 192,450 profiles) with the following rank percentages: 87.02\% newbies; 2.16\% Jr. Members; 4.47\% Members; 2.91\% Full Members; 1.27\% Hero Members; 0.38\% Legendary Members; 1.17\% Cooper Members; and, a really insignificant number of Donator/VIP/Administrator ranks (four in total). When considering that users are automatically labelled as Brand New on registration and they turn into Newbie on publishing one post, there is a significantly high proportion of basic ranks. Among paying members (Cooper, Donator, VIP), Cooper is, by far, the largest category among the paying members. At the end of the year of study, paying members produce revenue of  402,835 US Dollars (bitcoin equivalent), which remark the interest of Bitcointalk forum.

\begin{figure}[H] %\centering
\includegraphics[width=\columnwidth]{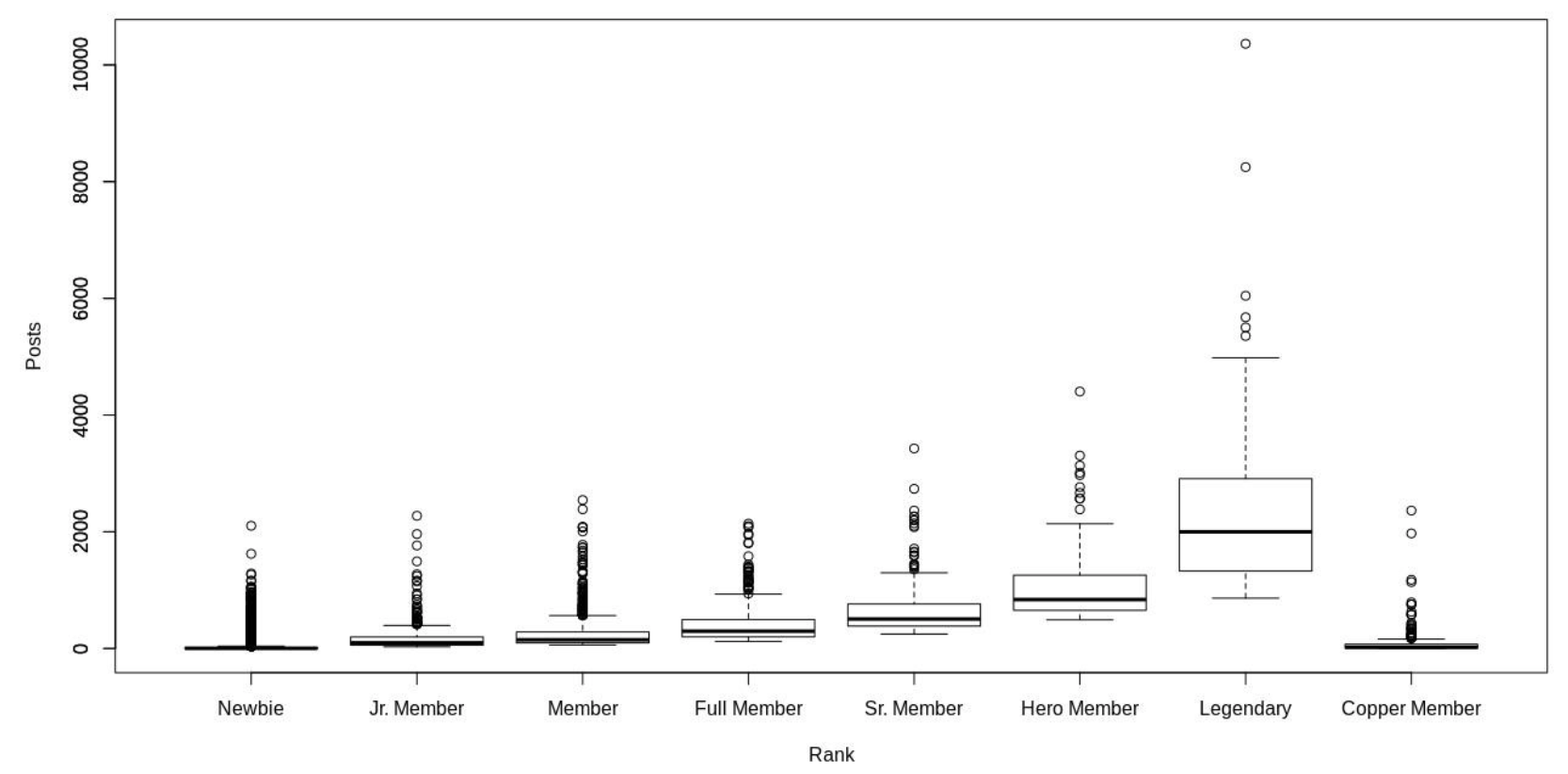} \caption{Number of posts per rank (boxplot).} \label{fig:Number-posts-per-rank-boxplot} \end{figure}

Profiles in TALK-DS  published a total of 858,270 posts during 2018, with a distribution of number of post per ranks in box-plot (see Figure~\ref{fig:Number-posts-per-rank-boxplot}; Donator, VIP, and Administrator are not taken into consideration, due to their lack of significance): the number of posts increase with the rank; additionally, the variation in number of post, which is a normal consequence from the tougher conditions required to get to the next position as the users go up in rank and it creates more diversity in the group of users. It is important to notice that the median is always situated in the lower half of the box. This means that the most usual status for users to be is closer to the transition to their current rank than the conditions to reach the next one. This idea is also supported by the position of the box: it is always situated very close to the lower whisker, an asymmetry that indicates that 50\% of the data in each rank is located there. Despite this strong asymmetry, some users rise above the rest with a large number of posts published (outliers). For instance, there is a significant group of Newbie users that have published more posts than some Legendary ones, but are stuck in their rank. This proves that posting a lot is not a synonym of achieving greater prestige in the community.

\begin{figure}[H] %\centering
	\begin{tabular}{ll}
 \includegraphics[width=0.485\textwidth]{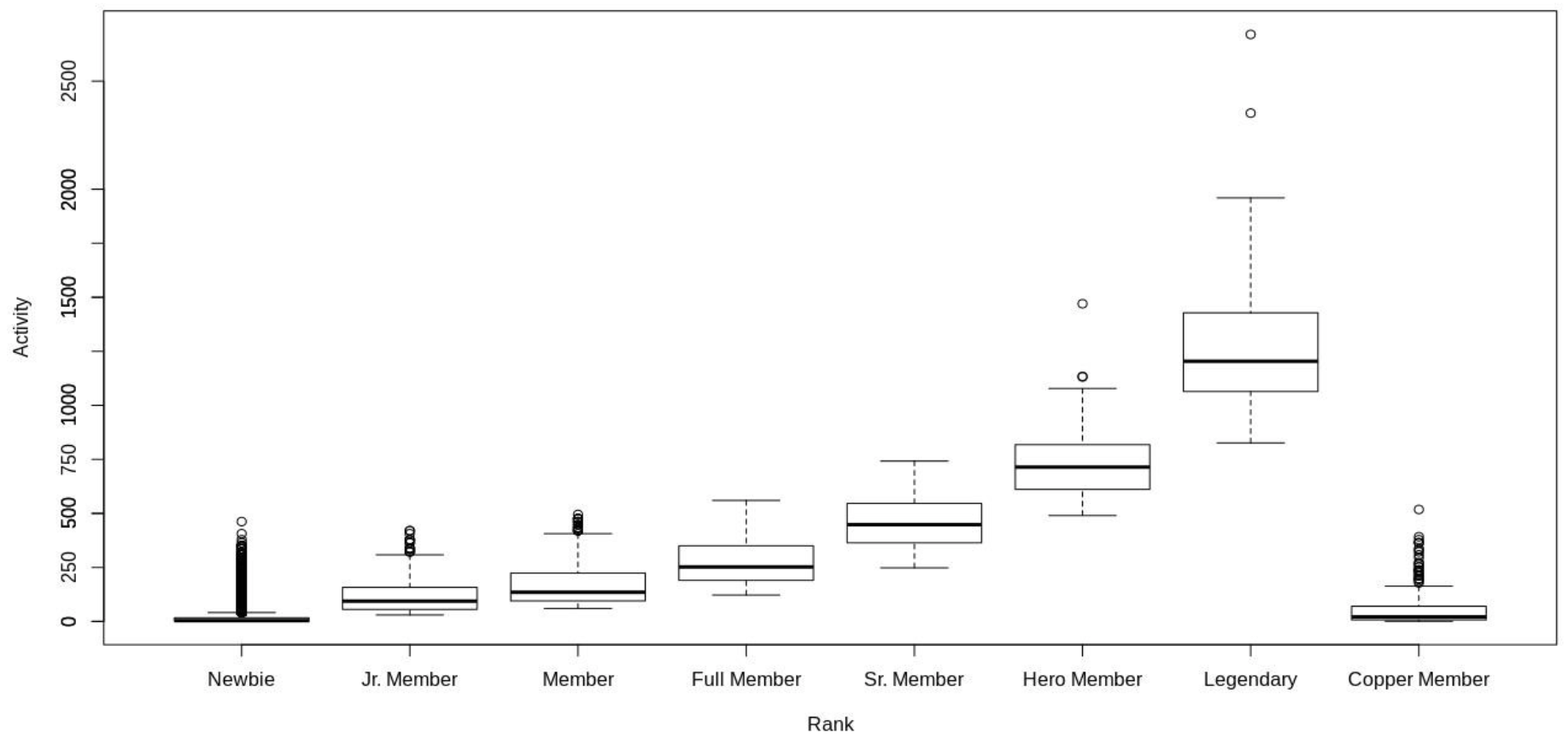} &  \includegraphics[width=0.485\textwidth]{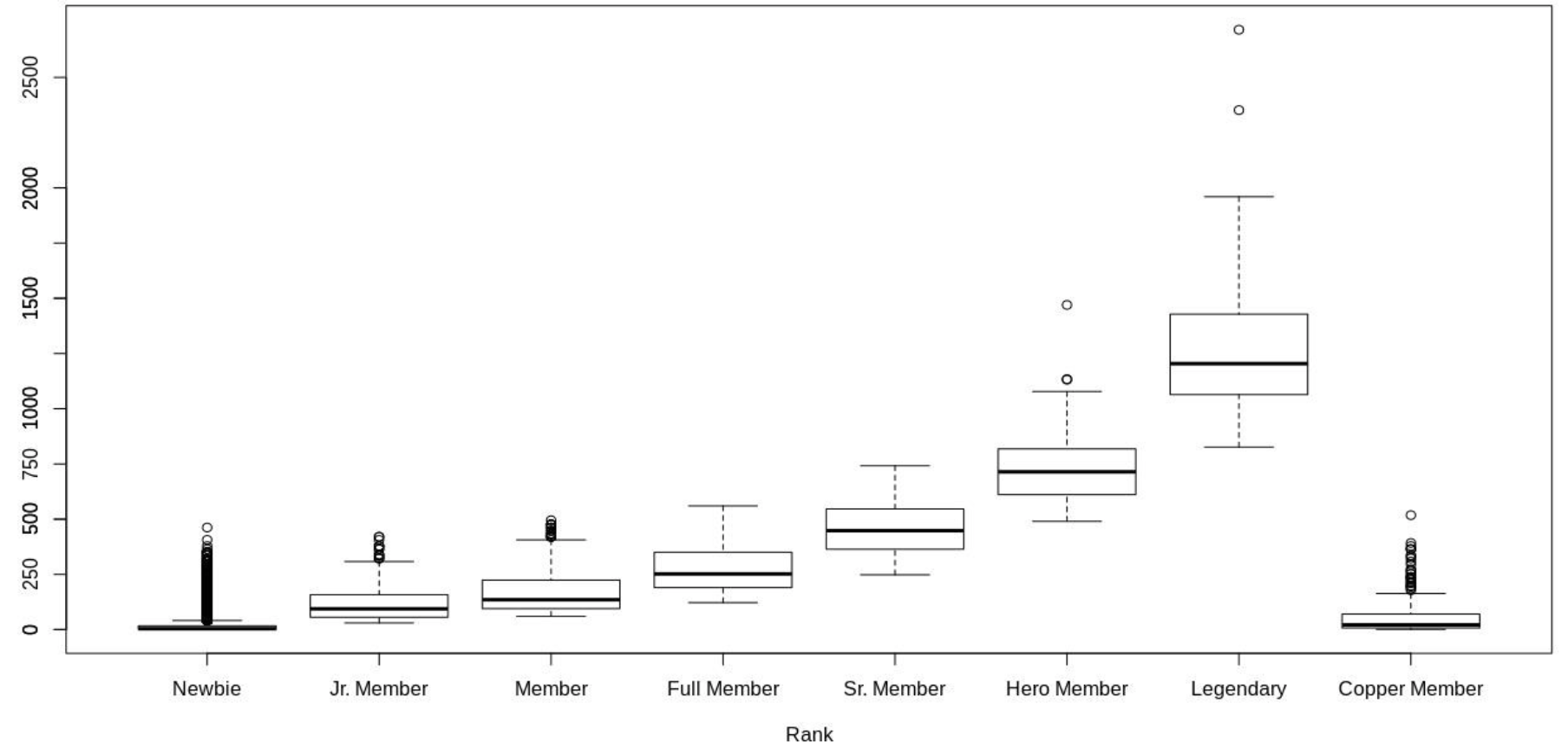} \\
\end{tabular}  \caption{Distribution of activity and merit per rank (box-plots).\label{fig:DistributionMeritPerRank}} 
\end{figure}

This can be further explained with the analysis of the two requirements to rank up: activity and merit (see Figure~\ref{fig:DistributionMeritPerRank}). A comparison between the activity box-plot and the posts one reveals a considerable softening in the asymmetry and the number of outliers. The reason is that the activity parameter is not affected by the number of posts that are published by users. Only 14 posts in two weeks are necessary to get the maximum of activity points. Rhythm and posting beyond that number are not considered. The Newbie is the only rank where there is almost no difference between activity and posts distribution. Thirty activity points are required to be a Jr. Member. Therefore, only one merit point is needed to rank up, but this is exactly what users that are represented in the outliers are lacking. Those Newbie users with hundreds of activity points would need further research. However, since they are not able to be merited for any single post, they are expected to be potential spammers. The size of boxes of the merit box-plot is very tiny, which means little variation among users according to their merit points. In this case, the median matches the number of points required to get to the next rank. The outliers that are on the limit required to change status of higher represent users with the potential to rank up. They have published fewer posts than required to change status, but they were welcomed by the community that recognizes them. Except for Hero members, they will eventually rank up automatically if they keep posting.

\begin{figure}[H] %\centering
 \includegraphics[width=0.8\textwidth]{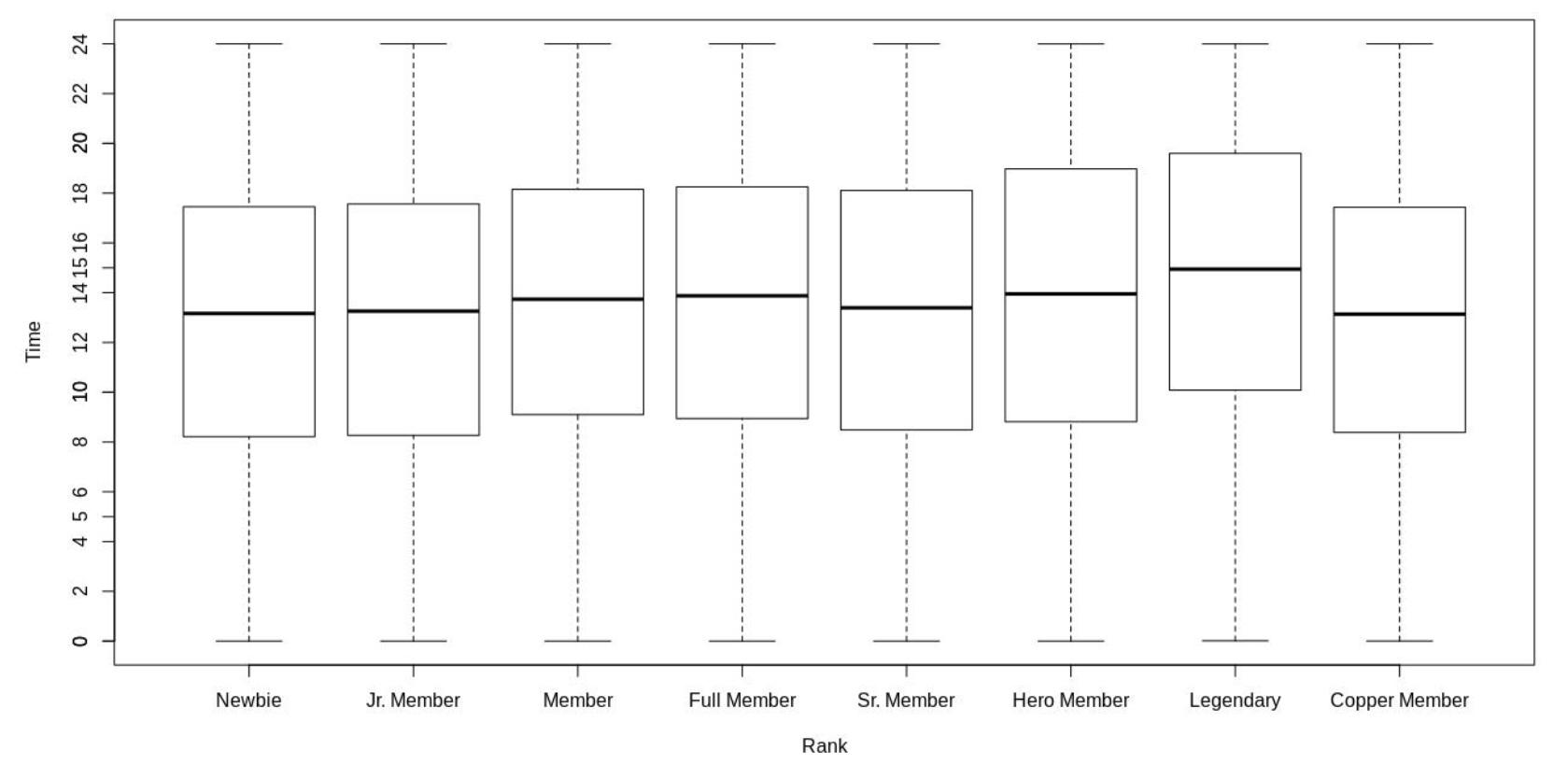}  %\\
 \caption{The influence of time of posts. \label{fig:InfluenceTimePosts}} 
\end{figure}

It is also interesting for this research to analyze how often users posted during 2018, as well as at what time of the day and moment of the year users participated the most in the forum. It is important to keep in mind that users post from different parts of the world, but Bitcointalk uses GMT+0 as local time. The box-plot shows that, in general, the time with the most activity is 13:26 and the most active period of the day is between 08:33 and 17:48. Newbie users strongly influenced the dataset, who represent 87\% of the data, and the possibility that the different ranks may have different schedules must be taken into consideration. The box-plot presented in Figure~\ref{fig:InfluenceTimePosts} shows this assumption.

The diagram confirms that there is no big difference between schedules, as the boxes in all cases stay approximately in the previously calculated range. The rank that differs the most is Legendary: the box is located between 10:05 and 19:36 and the median is situated at 14:57. This box plot confirms that no pattern of activity depends on the rank and, therefore, all of the users post in the same time slot.

\begin{figure}[H] %\centering
 \includegraphics[width=0.8\textwidth]{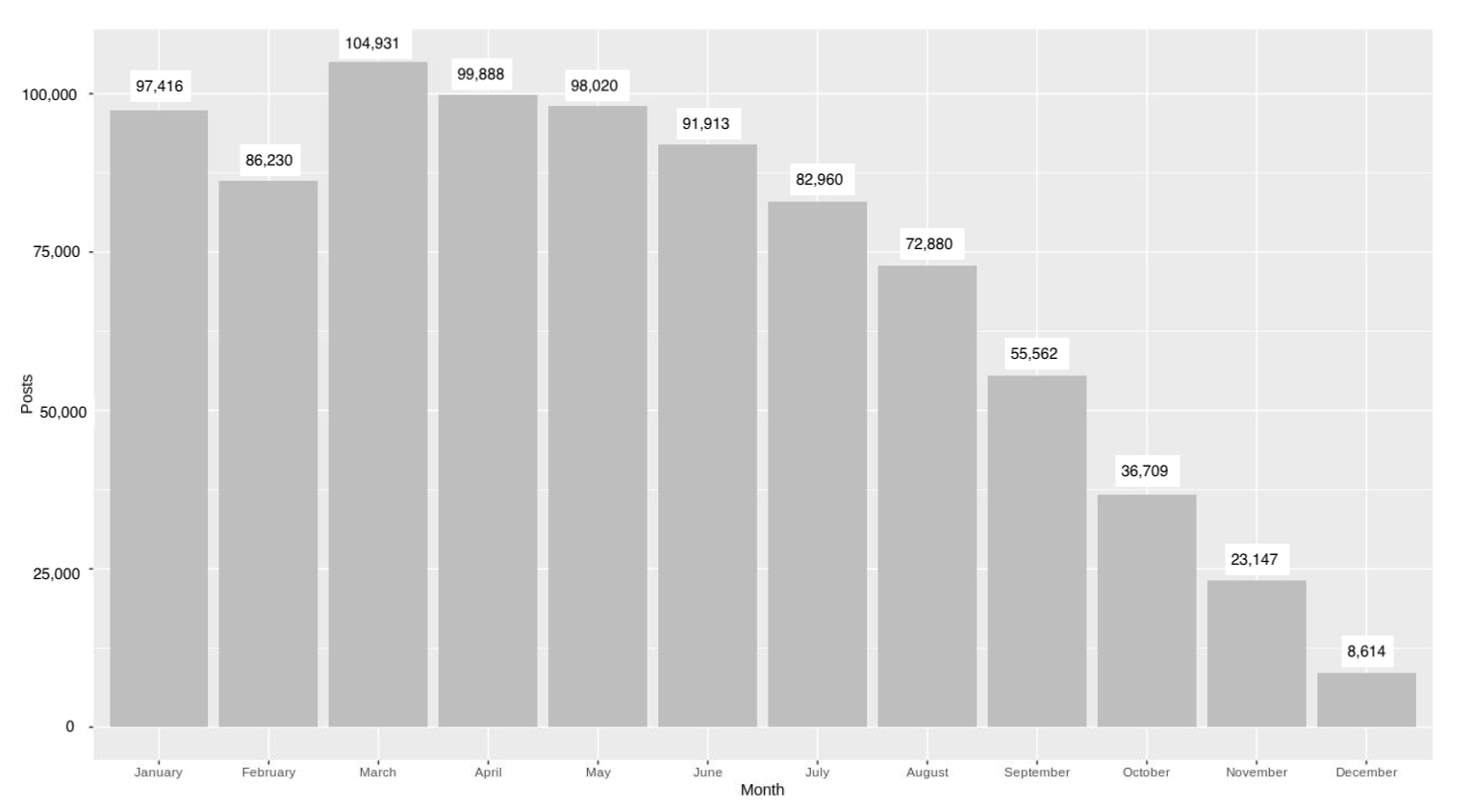}
 \caption{The influence of the time of the year. \label{fig:HistogramTimeYear}} 
\end{figure}

\begin{figure}[H] %\centering
 \includegraphics[width=0.8\textwidth]{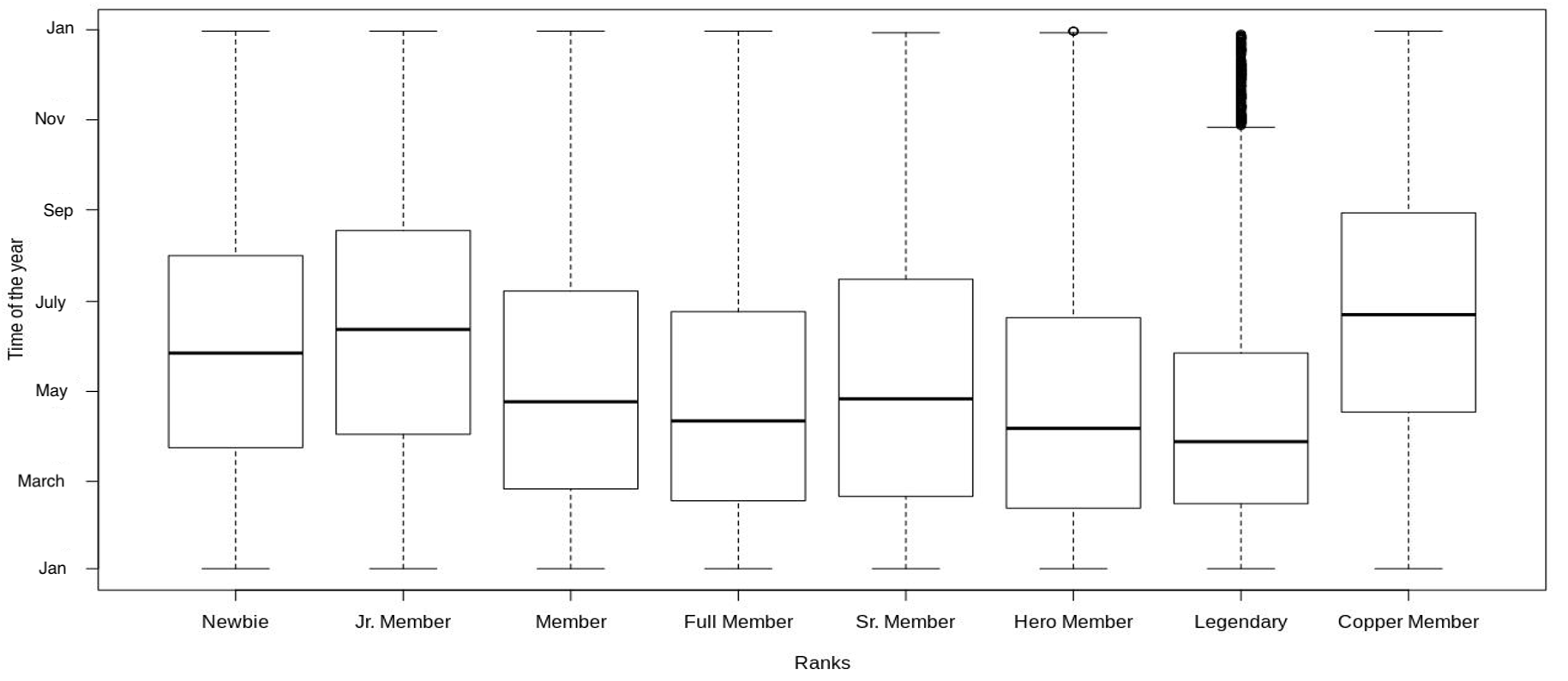} %\\
 \caption{Posts by time of the year and rank. \label{fig:PostsTimeYearRank}} 
\end{figure}

The histogram presented in Figure~\ref{fig:HistogramTimeYear} shows the distribution of posts according to the month they were posted. It shows the tendency of the forum clearly.  The users of the sample started 2018 posting 97,416, but this value decreased slightly the next month. It reached the maximum in March, and then it started to decrease progressively. From that month on, the decreases became more significant until December, when only 8614 posts were published. This fact reflects the loss of user participation in the forum. Because this histogram is strongly affected by the behavior of newbie users, the distribution of posts according to ranks was also checked and represented in the box-plot in Figure~\ref{fig:PostsTimeYearRank}.

The higher ranked positions present a higher asymmetry, with the exception of Jr. Member and Copper Member. The rank with the highest asymmetry is Legendary. Its box is located between 2nd  February and 27th May and the median is 28th  March. Another remarkable difference is the fact that its superior whisker is not located at the end of December, like the rest. It is situated on 27th October and the values beyond that day are considered outliers. This confirms that Legendary users have a more intense tendency; their posting activity starts to decrease from the end of March and it is almost non-existent in November and December. The graphics demonstrate the trajectory of the forum in 2018: it is intense at the beginning of the year and it declines over time, especially for the ranks with more weight in the forum.

\section{Results}
\label{sec:results}

\begin{figure}[H] %\centering
 \includegraphics[width=0.8\textwidth]{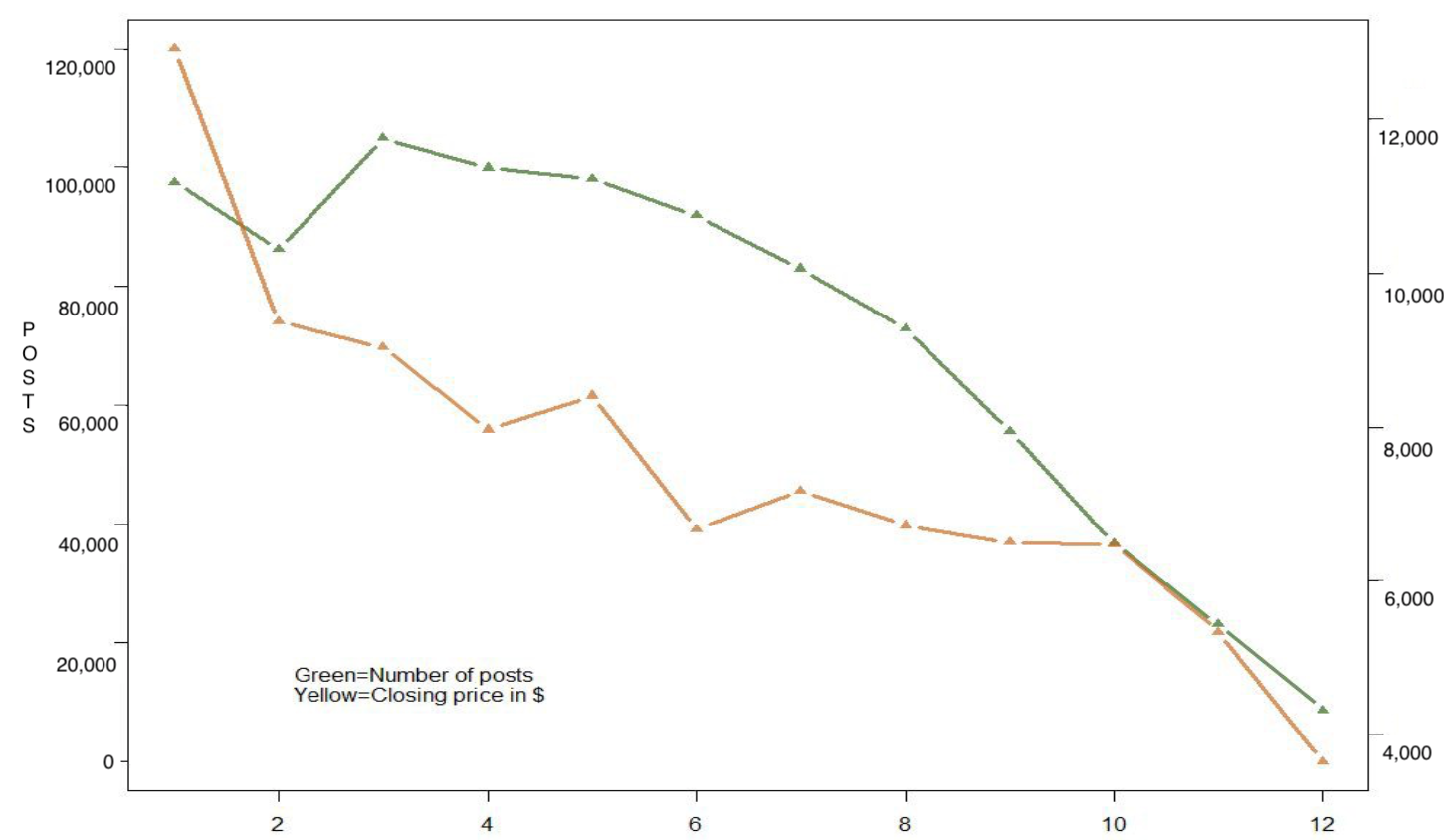} %\\
 \caption{Nominal comparison between the monthly number of posts and the closing value in 2018. \label{fig:NominaComparisonMonthlyPostsClosingValue}} 
\end{figure}

\begin{figure}[H]%\centering
 \includegraphics[width=0.8\textwidth]{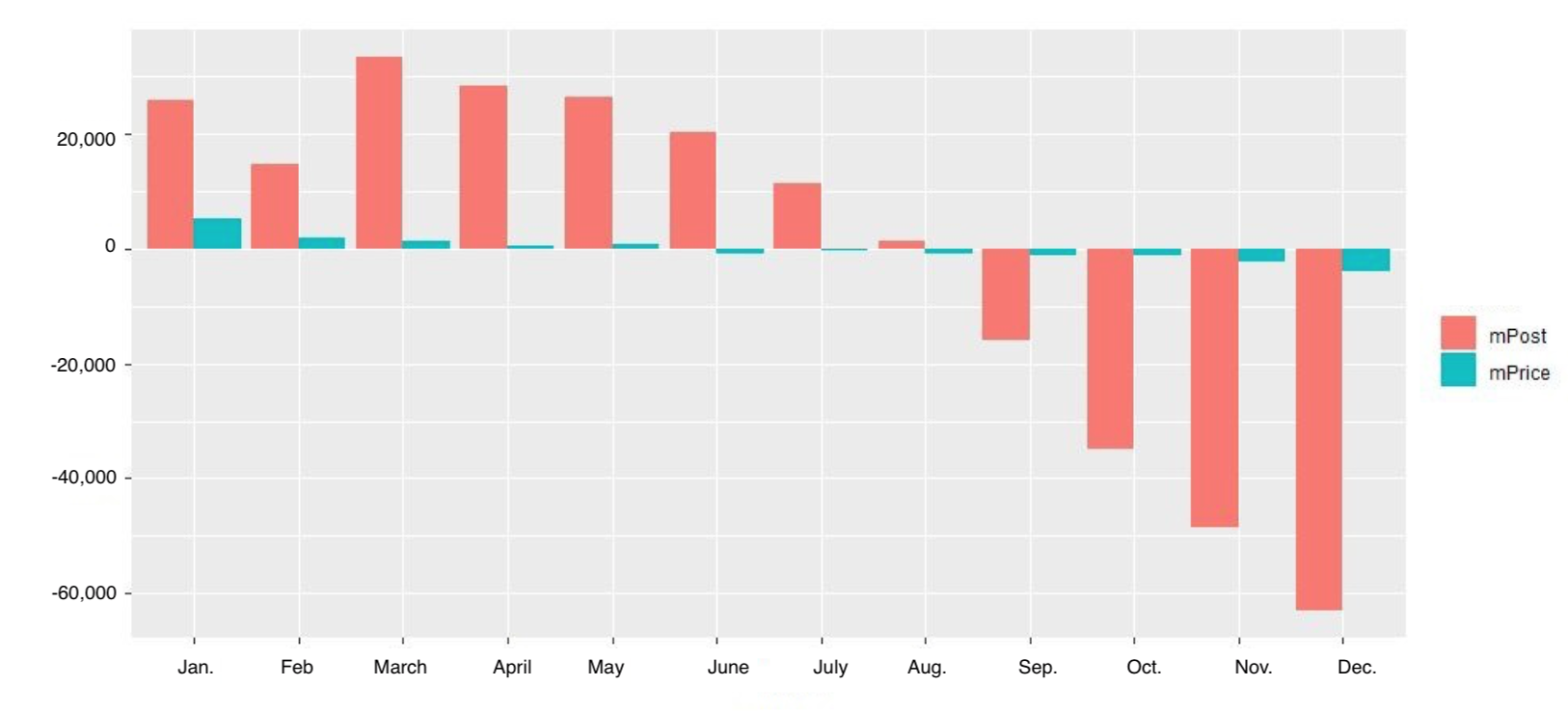} %\\
 \caption{Variation of the monthly posts and the mean value of Bitcoin in 2018. \label{fig:VariationMonthlyPostsMeanValue}} 
\end{figure}

In order to assess the credibility and reliability of the forum, the evolution of its activity is visualized  along with the Pearson correlation for daily increments and a proper interpretation. The time scale for the analysis of the data set is set to one day as well as one month. For a monthly analysis, Figure~\ref{fig:NominaComparisonMonthlyPostsClosingValue} shows the nominal comparison between monthly posts and closing price of BTC in USD (\$) during 2018. Additionally, Figure~\ref{fig:VariationMonthlyPostsMeanValue} shows the connection between the variation in posts per month and the monthly value of BTC in 2018.

\unskip

If we correlate the data between monthly posts and the monthly closing price of BTC, we have a result of 0.756, which shows the coincidences between the activity and the value of Bitcoin. In short, August was the turning point in the relationship between the value of the cryptocurrency and the activity in the forum. In September and October, BTC kept its value but the forum had less activity. As the value decreased constantly in those months, more people started to be less active, because there was no information in the market to provoke a reaction in the forum. Most of the users ceased their activity in the forum, because they consider that had been a hit, but had started to lose its prestige and was normalized. This is consistent with the statement that was made in the first pages of this paper: the value of cryptocurrencies can suffer many changes.

\begin{figure}[H] %\centering
 \includegraphics[width=0.8\textwidth]{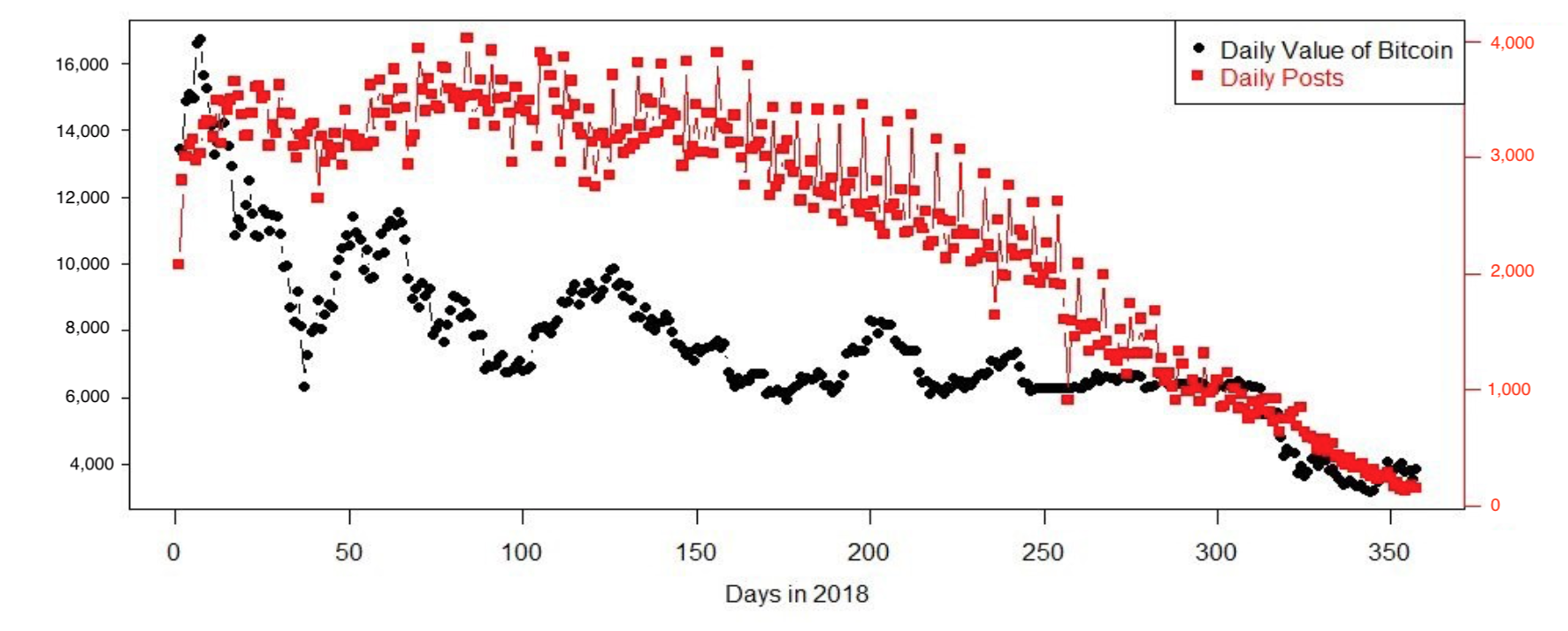} %\\
 \caption{Nominal comparison between the daily number of posts and the closing value in 2018. \label{fig:NominaComparisonDailyPostsClosingValue}} 
\end{figure}

\begin{figure}[H] %\centering
 \includegraphics[width=0.8\textwidth]{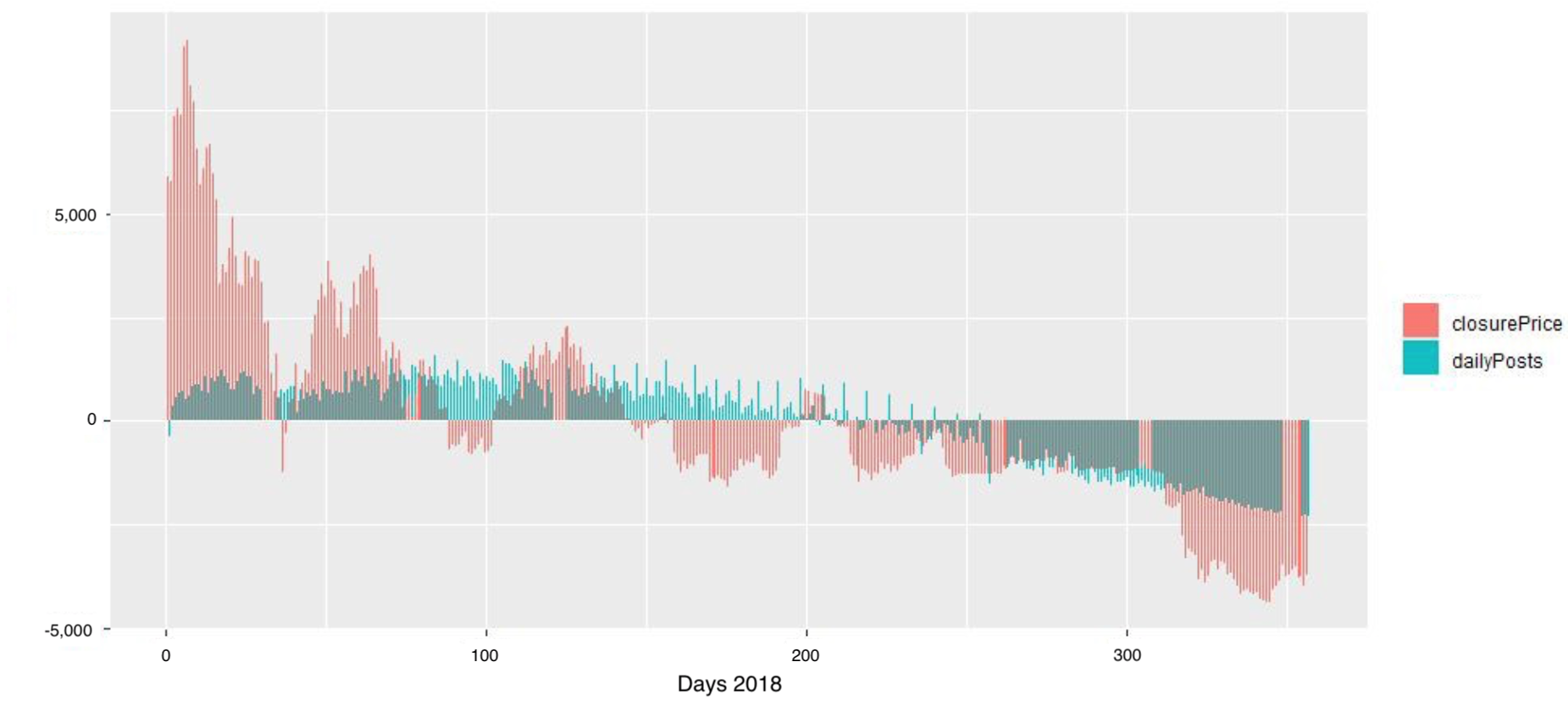} %\\
 \caption{Comparison between the variation of posts per day and the daily closing value in 2018. \label{fig:ComparisonVariationPostsPerDayClosingValue}} 
\end{figure}

Figure~\ref{fig:NominaComparisonDailyPostsClosingValue} shows the nominal comparison between daily posts and closing value of BTC in USD; and, Figure~\ref{fig:ComparisonVariationPostsPerDayClosingValue} shows a comparison between the variation of posts per day and the daily closing value of BTC during the year 2018. The result of the correlation is 0.672. It is lower than the first one, because, in the previous one, the collection of data per month cancels certain values that are present here. Therefore, regarding the possible relationship between the fluctuations in user activity and the closing price, the connection can be demonstrated quite easily with the high activity level in the forum in the first 100 days of the year, when the value of BTC was high and the value at the end of 2017 had already been promising. The critical point here is the period between days 200 and 210, where both variables change. The users of the forum interact, but their activity starts to decrease as the value starts to fall. The blue peaks between days 200 and 250 are interactions about exchanges and discussions about the economy and the fall of Bitcoin. From day 250 to the end, both variables decrease. Around day 350 the activity of the users seems to be in the mean. However, this looks like a one-time exception to the dynamic. That same exception happened on day 1, where the daily posts are lower than the mean.

\unskip 

There is another interesting dynamic in the cryptocurrencies market, named Pump-and-Dump~\cite{li2019cryptocurrency}. This phenomenon is a dynamic from a group of investors to increase the value of a currency at the time of sale with false statements. There is a study~\cite{PumpDumpSchemes} that reveals the existence of these groups and their activity between January and June this year, as the value had a great variation. During the second semester, where the value was more static, these groups stopped their dynamic. Some users with experience in the market stated in some posts that joining these groups was not advisable, because the profit goes to the leaders of the group and the risks/profit ratio is very high, especially if you are a new in the field of cryptocurrencies.

\begin{figure}[H]%\centering
 \includegraphics[width=0.8\textwidth]{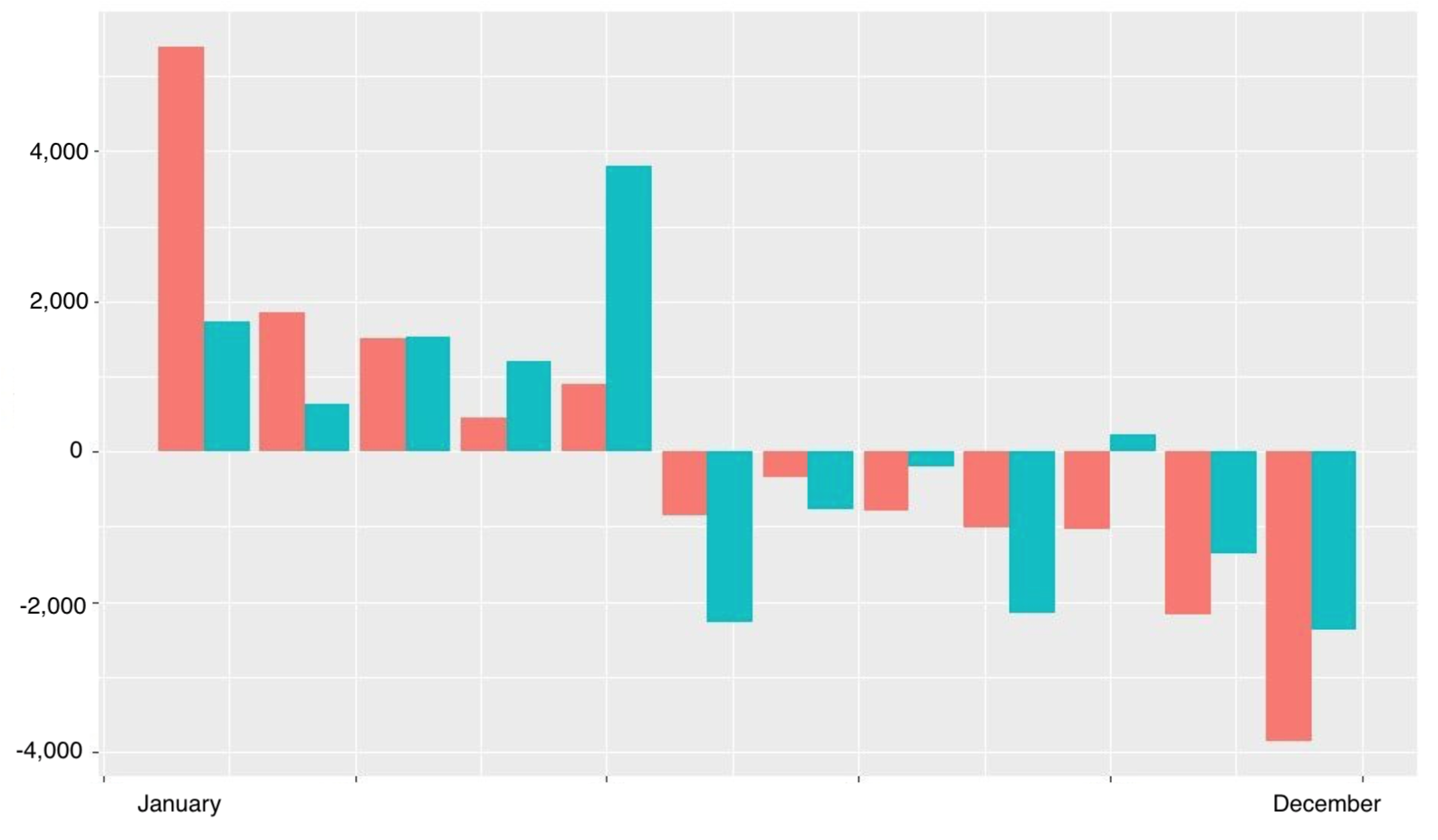} %\\
 \caption{Variation of the average of quotes per month and the closing value in 2018. Red: Price and Green: Quotes. \label{fig:VariationAverageQuotesMonthClosingValue}} 
\end{figure}

Before moving on to the next graphic, an introduction is required. In forums, posts from users are not the only source of information, the reactions from other users in a topic also provide data. The most common tool for publishing a reply or a response to a post is the quote. Most of the forums allow for posts containing the original contribution from another user. This way, quotes are useful for the first user, who knows that their posts have response or interactions with the community. Figure~\ref{fig:VariationAverageQuotesMonthClosingValue} compares the variation of the average value per month of quotes and the average of  the closing value, which reveals an abnormal reaction from the users to a decrease in the value of Bitcoin. In May and October, the quotes are quite relevant and they introduce a massive loss in value, as seen in June and November. The correlation between both data is 0.6864684. It confirms the statement that the system of quoting is a valid measure for locating the loss in value of this cryptocurrency.

\begin{figure}[H] %\centering
 \includegraphics[width=0.8\textwidth]{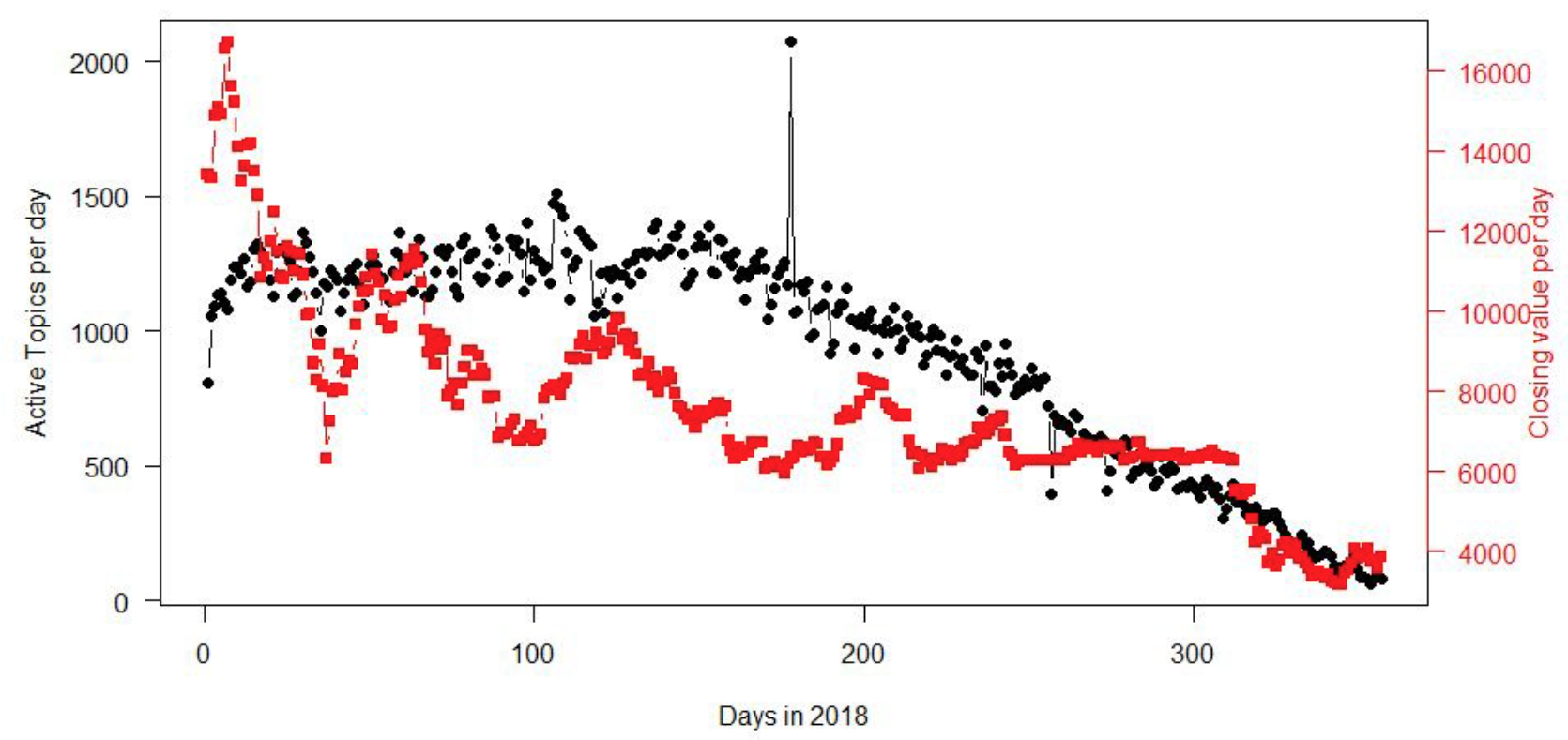} %\\
 \caption{Comparison between the daily active topics and the closing value in 2018. \label{fig:ComparisonDailyActiveTopicsClosingValue}} 
\end{figure}

Figure~\ref{fig:ComparisonDailyActiveTopicsClosingValue} shows the relationship between the active topics per day and the daily closing value during 2018. The information that can be extracted from this graphic is especially relevant with regard to the activity in the forum when the values are so extreme. In other words, the forum shows less activity in days when BTC has a rising value, as well as the days immediately afterwards than a day where the value falls. In that case, the activity in the forum increases. This behavior can be also seen in forums from other areas, like sports or video games. If there is an important announcement, the forum has more views in a close group of topics that are directly related to that announcement. However, when there is no event or relevant situation, there are more active topics. Moreover, the forum keeps the topics active while they are relevant. In other words, there will be always a correlation between the activity in a forum and the reason for the creation of the forum.

\begin{figure}[H]%\centering
 \includegraphics[width=0.8\textwidth]{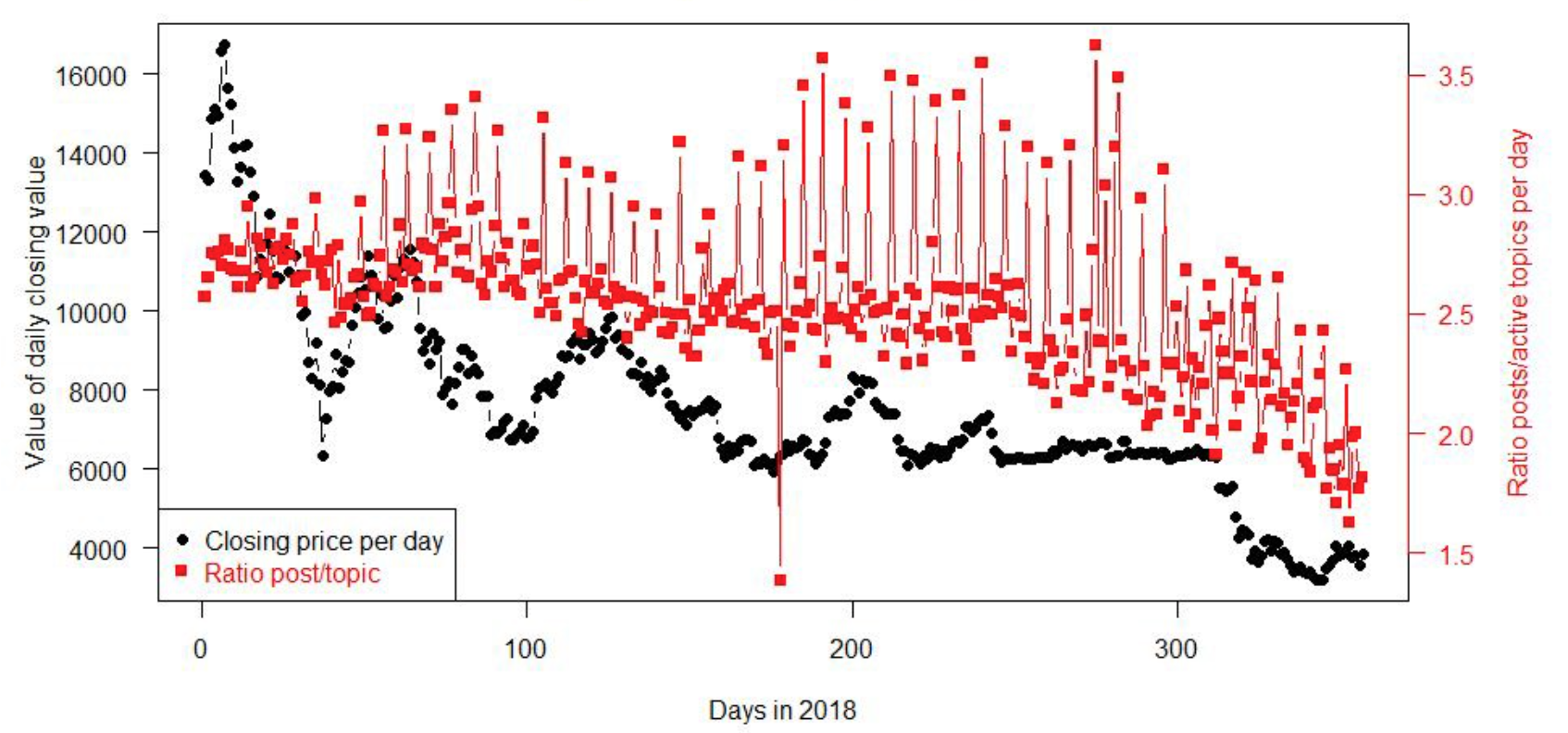} %\\
 \caption{Comparison between the post/topic ratio and the daily closing value in 2018 \label{fig:ComparisonPostTopiRatioDailyClosingValue}} 
\end{figure}

Figure~\ref{fig:ComparisonPostTopiRatioDailyClosingValue} compares the post/topic ratio and the closing price, as an extension of the idea that is exposed in the previous figure. This graphic is a more detailed representation of the activity of a community in a forum. To be more precise, it shows that the post/topic ratio increases regularly one day per week. This is very important for both the community and the purpose of the forum. For instance, in a video game forum, there is a maintenance operation or an update of the game each week that disables the game and the players to discuss the upcoming news or developing a hypothesis related to the new version. In our case, the same can be applied to weekly topics where there is an update related to Bitcoin. That update can be linked to different aspects, such as value, uses, news about its future, or its implementation. The most relevant idea extracted from this analysis is the correlation of these data. The result is 0.43, a small number when compared to the 0.65 coefficient that can be seen in Figure~\ref{fig:ComparisonDailyActiveTopicsClosingValue}. That result confirms that a forum will be active for as long as the topic that originated it is relevant. However, the lower result proves that, in regular periods, the activity focuses on a few topics and it is primarily composed of the reactions to the present economic value of BTC or estimations of it.

\unskip

\section{Predictive Power}
\label{sec:predictive:power}

\begin{figure}[H] 
 \includegraphics[width=0.8\textwidth]{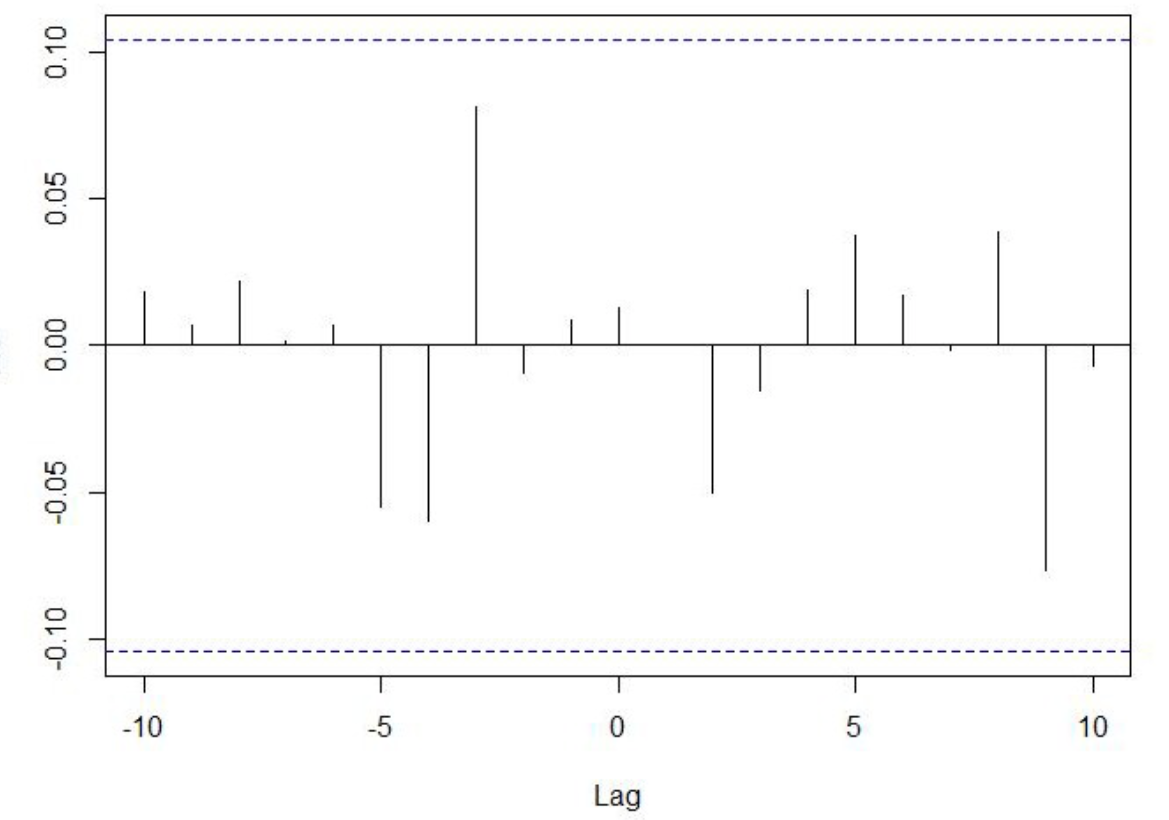} %\\
 \caption{Cross-correlation between variations in post/topic ratio and value. \label{fig:Cross-correlationVariationsPostTopicRatioValue}} 
\end{figure}

Coming back to the main objective of this research, and in order to give relevant information to cryptocurrency users, we inspect the potential of using forum data to predict the values of cryptocurrencies.  Figure~\ref{fig:Cross-correlationVariationsPostTopicRatioValue} shows the cross-correlation between variations in the post/topic ratio and closing value of BTC per day. A window of 10~days is used to observe a quick reaction from the users that can predict the value of Bitcoin. Table~\ref{tab:Lag-post/topic-ratio-value} introduces the nominal values of the graphic, as there can be doubts to determine the maximum value of the cross- correlation.

\begin{table}[H]
%\tablesize{\scriptsize}
\caption{Values in the lag window (variations in post/topic ratio and value). \label{tab:Lag-post/topic-ratio-value}}
%\setlength{\cellWidtha}{\columnwidth/10-2\tabcolsep+0.0in}
%\setlength{\cellWidthb}{\columnwidth/10-2\tabcolsep+0.0in}
%\setlength{\cellWidthc}{\columnwidth/10-2\tabcolsep-0.0in}
%\setlength{\cellWidthd}{\columnwidth/10-2\tabcolsep-0.0in}
%\setlength{\cellWidthe}{\columnwidth/10-2\tabcolsep-0.0in}
%\setlength{\cellWidthf}{\columnwidth/10-2\tabcolsep-0.0in}
%%\setlength{\cellWidthg}{\columnwidth/10-2\tabcolsep-0in}
%\setlength{\cellWidthh}{\columnwidth/10-2\tabcolsep-0in}
%\setlength{\cellWidthi}{\columnwidth/10-2\tabcolsep-0.0in}
%\setlength{\cellWidthj}{\columnwidth/10-2\tabcolsep-0in}
%\scalebox{1}[1]{\begin{tabularx}{\columnwidth}
%{>{\PreserveBackslash\centering}p{\cellWidtha}
%>{\PreserveBackslash\centering}p{\cellWidthb}
%{\PreserveBackslash\centering}p{\cellWidthc}
%>{\PreserveBackslash\centering}p{\cellWidthd}
%{\PreserveBackslash\centering}m{\cellWidthe}
%>{\PreserveBackslash\centering}m{\cellWidthf}
%>{\PreserveBackslash\centering}m{\cellWidthg}
%>{\PreserveBackslash\centering}m{\cellWidthh}
%{\PreserveBackslash\centering}m{\cellWidthi}
%>{\PreserveBackslash\centering}m{\cellWidthj}}
%\toprule
\begin{tabular}{llllllllll}
\multicolumn{10}{c}{\textbf{Lag/Cross-Correlation}}\\ 
  $-$10	& $-$9	& $-$8	& $-$7	& $-$6	& $-$5	& $-$4	& $-$3	& $-$2	& $-$1	\\
 0.018& 	0.007	& 0.022	& 0.002	& 0.007	 & $-$0.055	& $-$0.059 & 	0.081 &	$-$0.009	& 0.009	\\ 
 %  0 &&&&&&&&&\\
 % 0.013 &&&&&&&&&\\ \hline
  10 & 	9&	8	&7	&6	&  5	&4	&3	&2	&1	\\
$-$0.007 &	$-$0.076 &	0.038 &	$-$0.001 &	0.017	& 0.038	 & 0.019	& $-$0.015	& $-$0.050& 	0.000	\\

\end{tabular}
\end{table}

\begin{figure}[H]
 \includegraphics[width=0.8\textwidth]{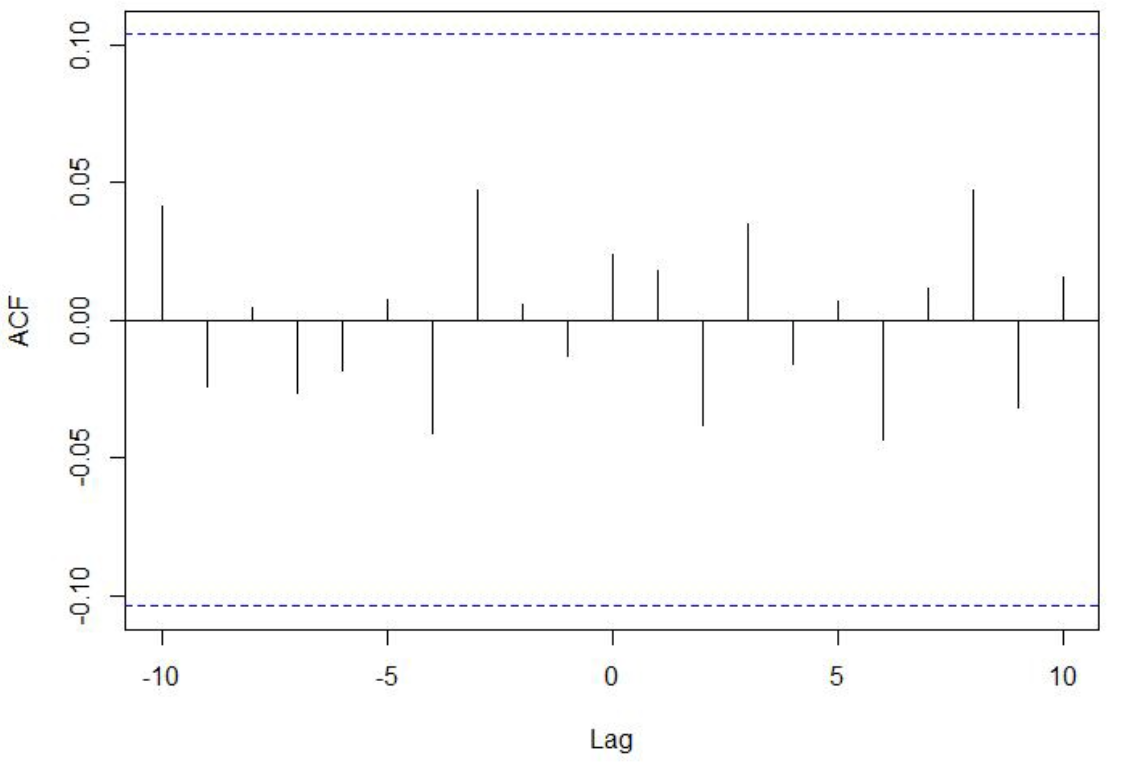} %\\
 \caption{Cross-correlation between variations in daily posts and value. \label{fig:Cross-correlationVariationsDailyPostsValue}} 
\end{figure}

Figure~\ref{fig:Cross-correlationVariationsDailyPostsValue} shows the highest value in h = $-$3. This means that an uprising value in the post/topic ratio can be directly linked to the closing value in the following three days. This ratio can be used as evidence of the concentration of the activity in a small set of topics---as explained above---despite the low correlation between the ratio and closing value. 

The cross-correlation with the posts and active topics was also considered. Regarding posts, the results peak in h = $-$3, so this shows the period when the community can predict a change, like the post/topic ratio. However, the figure also represents the second-highest value, in h~=~8. In this case, it can show a weekly activity in different topics, as the value in the previous graphic is also significant in h~=~8. Table~\ref{tab:Lag-variations-dayly-pots-value} shows the value of each lag period.

\begin{table}[H]
%\tablesize{\scriptsize}
\caption{Values in the lag window (variations in daily posts and value). \label{tab:Lag-variations-dayly-pots-value}}
%\setlength{\cellWidtha}{\columnwidth/10-2\tabcolsep+0.0in}
%\setlength{\cellWidthb}{\columnwidth/10-2\tabcolsep+0.0in}
%\setlength{\cellWidthc}{\columnwidth/10-2\tabcolsep-0.0in}
%\setlength{\cellWidthd}{\columnwidth/10-2\tabcolsep-0.0in}
%\setlength{\cellWidthe}{\columnwidth/10-2\tabcolsep-0.0in}
%\setlength{\cellWidthf}{\columnwidth/10-2\tabcolsep-0.0in}
%\setlength{\cellWidthg}{\columnwidth/10-2\tabcolsep-0in}
%\setlength{\cellWidthh}{\columnwidth/10-2\tabcolsep-0in}
%\setlength{\cellWidthi}{\columnwidth/10-2\tabcolsep-0.0in}
%\setlength{\cellWidthj}{\columnwidth/10-2\tabcolsep-0in}
%\scalebox{1}[1]{\begin{tabularx}{\columnwidth}
%{>{\PreserveBackslash\centering}p{\cellWidtha}
%>{\PreserveBackslash\centering}p{\cellWidthb}
%>{\PreserveBackslash\centering}p{\cellWidthc}
%>{\PreserveBackslash\centering}p{\cellWidthd}
%>{\PreserveBackslash\centering}m{\cellWidthe}
%>{\PreserveBackslash\centering}m{\cellWidthf}
%>{\PreserveBackslash\centering}m{\cellWidthg}
%>{\PreserveBackslash\centering}m{\cellWidthh}
%>{\PreserveBackslash\centering}m{\cellWidthi}
%>{\PreserveBackslash\centering}m{\cellWidthj}}
%\toprule

\begin{tabular}{llllllllll}

\multicolumn{10}{c}{\textbf{Lag/Cross-Correlation}}\\ 
 $-$10	& $-$9	& $-$8	& $-$7	& $-$6 & $-$5	& $-$4	& $-$3	& $-$2	& $-$1	\\
 0.041	&$-$0.024	&0.004	&$-$0.026	&$-$0.018 & 0.008&
 $-$0.041	&0.047	&0.006	&$-$0.013 \\ 

%Lag &  0 &&&&\\
% Cross-correlation & 0.024 &&&&\\ \hline
 10 & 	9&	8	&7	&6	&5	&4	&3	&2	&1	\\
0.016&	$-$0.032	&0.047	&$-$0.012	&$-$0.043  & $-$0.007	&$-$0.016	&0.035	&$-$0.038	&0.018		\\

\end{tabular}
\end{table}

\nointerlineskip

\begin{figure}[H]
 \includegraphics[width=0.8\textwidth]{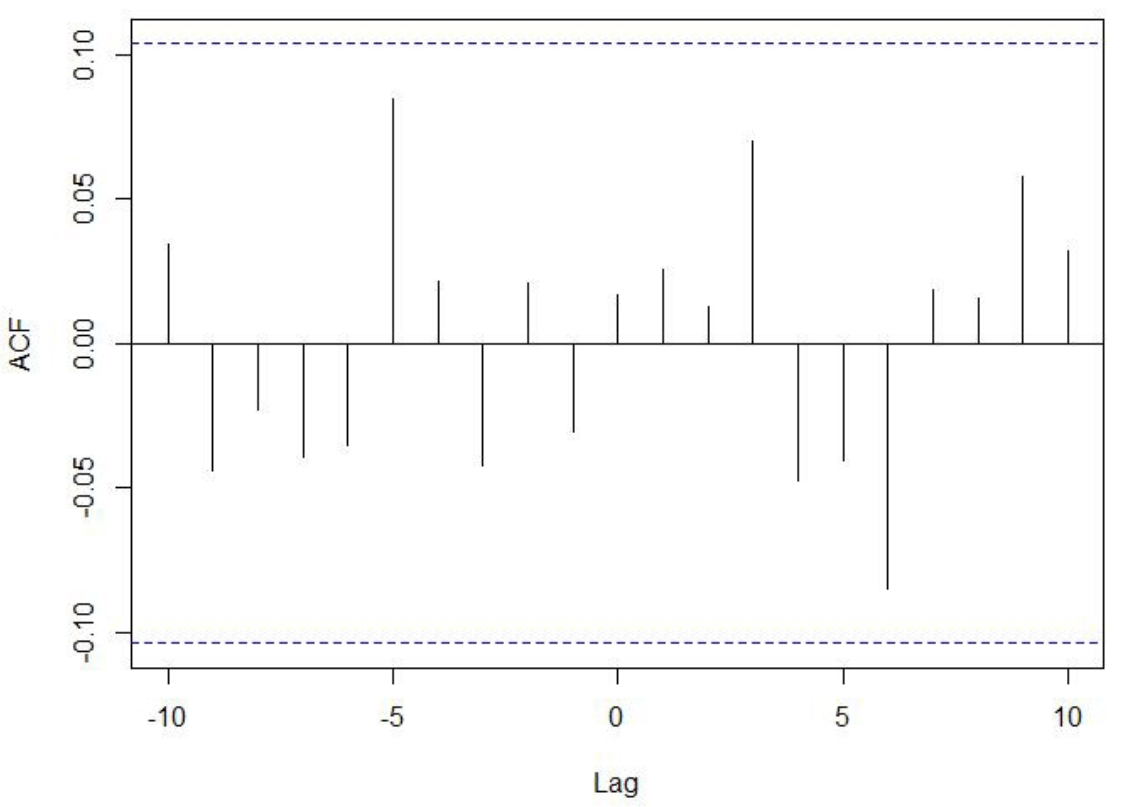} %\\
 \caption{Cross-correlation between variations in daily topics and value \label{fig:Cross-correlationVariationsDailyTopicsValue}} 
\end{figure}

On the other hand, the cross-correlation with the topics shows different results in the positive values. However, it must be noted that a positive cross-correlation in Figure~\ref{fig:Cross-correlationVariationsDailyTopicsValue} indicates a huge number of topics being open. The critical points here are the negative values, which show an inverse relationship and can cause a reduction in the number of active~topics.

Here (Table~\ref{tab:Lag-variations-dayly-topics-value}), the value in the lag period with the highest impact---represented at h = $-$3---is negative. This is relevan, because the data from the last three graphics show that the community produces a high number of posts about a reduced set of topics three days before a change in the value of Bitcoin. Besides, the most negative value, in h = 6, can be considered as a reaction to the value of Bitcoin, as evidenced by the positive number of posts in the following days. The positive values, however, are also interesting. In this case, the value in h = $-$5 is the peak in this cross-correlation. The activity seems to be minimal and it proves, once again, that there are more active topics in the forum, since there is no significant prediction or feedback about BTC and its value. 
\begin{table}[H]
%\tablesize{\scriptsize}
\caption{Values in the lag window (variations in daily topics and value). \label{tab:Lag-variations-dayly-topics-value}}
%\setlength{\cellWidtha}{\columnwidth/10-2\tabcolsep+0.0in}
%\setlength{\cellWidthb}{\columnwidth/10-2\tabcolsep+0.0in}
%\setlength{\cellWidthc}{\columnwidth/10-2\tabcolsep-0.0in}
%\setlength{\cellWidthd}{\columnwidth/10-2\tabcolsep-0.0in}
%\setlength{\cellWidthe}{\columnwidth/10-2\tabcolsep-0.0in}
%\setlength{\cellWidthf}{\columnwidth/10-2\tabcolsep-0.0in}
%\setlength{\cellWidthg}{\columnwidth/10-2\tabcolsep-0in}
%\setlength{\cellWidthh}{\columnwidth/10-2\tabcolsep-0in}
%\setlength{\cellWidthi}{\columnwidth/10-2\tabcolsep-0.0in}
%\setlength{\cellWidthj}{\columnwidth/10-2\tabcolsep-0in}
%\scalebox{1}[1]{\begin{tabularx}{\columnwidth}
%{>{\PreserveBackslash\centering}p{\cellWidtha}
%>{\PreserveBackslash\centering}p{\cellWidthb}
%>{\PreserveBackslash\centering}p{\cellWidthc}
%>{\PreserveBackslash\centering}p{\cellWidthd}
%>{\PreserveBackslash\centering}m{\cellWidthe}
%>{\PreserveBackslash\centering}m{\cellWidthf}
%>{\PreserveBackslash\centering}m{\cellWidthg}
%>{\PreserveBackslash\centering}m{\cellWidthh}
%>{\PreserveBackslash\centering}m{\cellWidthi}
%>{\PreserveBackslash\centering}m{\cellWidthj}}
%\toprule
\begin{tabular}{llllllllll}
\multicolumn{10}{c}{\textbf{Lag/Cross-Correlation}}\\ 
 $-$10	& $-$9	& $-$8	& $-$7	& $-$6  & $-$5	& $-$4	& $-$3	& $-$2	& $-$1	\\
 0.034&	$-$0.044	&$-$0.023	&$-$0.039	&$-$0.035 & 0.085	& 0.022	& $-$0.042	& 0.021	& $-$0.030\\ 

%Lag &  0 &&&&\\
% Cross-correlation & 0.017 &&&&\\ \hline
10 & 	9&	8	&7	&6 &	5	&4	&3	&2	&1	\\
0.032	& 0.058	& 0.016	& $-$0.019	& $-$0.085 & $-$0.040	&-0.047	&0.070	&0.013	&0.026	\\

\end{tabular}
\end{table}

From the last three cross-correlation diagrams, we can conclude that the forum can predict the value of BTC in a period of three days.

\section{Discussion}
\label{sec:discussion}

Previous research~\cite{10.1371/journal.pone.0161197} proved a prediction of BTC values derived from Bitcointalk activity. That study, which was presented in 2016, considered the whole lifetime of the forum and its posts. Nevertheless, our society nowadays has more information and it can access a high amount of data, so analyzing all of the related topics will be interesting to understand the relationship between the evolution of a digital asset and its community of users. The results of that study confirm that the direct relationship between posts and the price of BTC were close to the values that were obtained in  this research. The amount of information nowadays as compared to years ago confirms that society has more information available; therefore, the implementation of Machine Learning with the new data and the evolution of the community would provide a better result than the one of the study carried out in 2016. The historical peak at the end of 2017 and the later fall during 2018 is a relevant event in the life of the forum, which deserves a new analysis. The activity of a large number of users, the continuous visits to the websites related to the BTC, and the interactions between them are a very precise measure of the evolution of Bitcoin. Additionally, in~\cite{10.3389/fbloc.2020.00001}, the authors introduce a methodology for detecting shifts in price trends (bitcoin and Ethereum) that are based on mono-phase and multi-phase analysis, these shifts can be considered price outliers.  The study uses reddit as observatory data and, being reddit a general-purpose social network, specific subreddits are selected, and natural language processing (NLP) is applied to filter content relevant to shifts in price trends. With Bitcointalk, an specific financial forum, we can moreover consider users’ characteristics, like merit, position, etc. Nonetheless, the work shown in~\cite{10.3389/fbloc.2020.00001} focuses on location of potential causes for shits. In this respect, similar NLP could also be applied in our methodology to provide supplementary descriptive information along with the detected outliers. 

\begin{figure}[H]
 \includegraphics[width=0.8\textwidth]{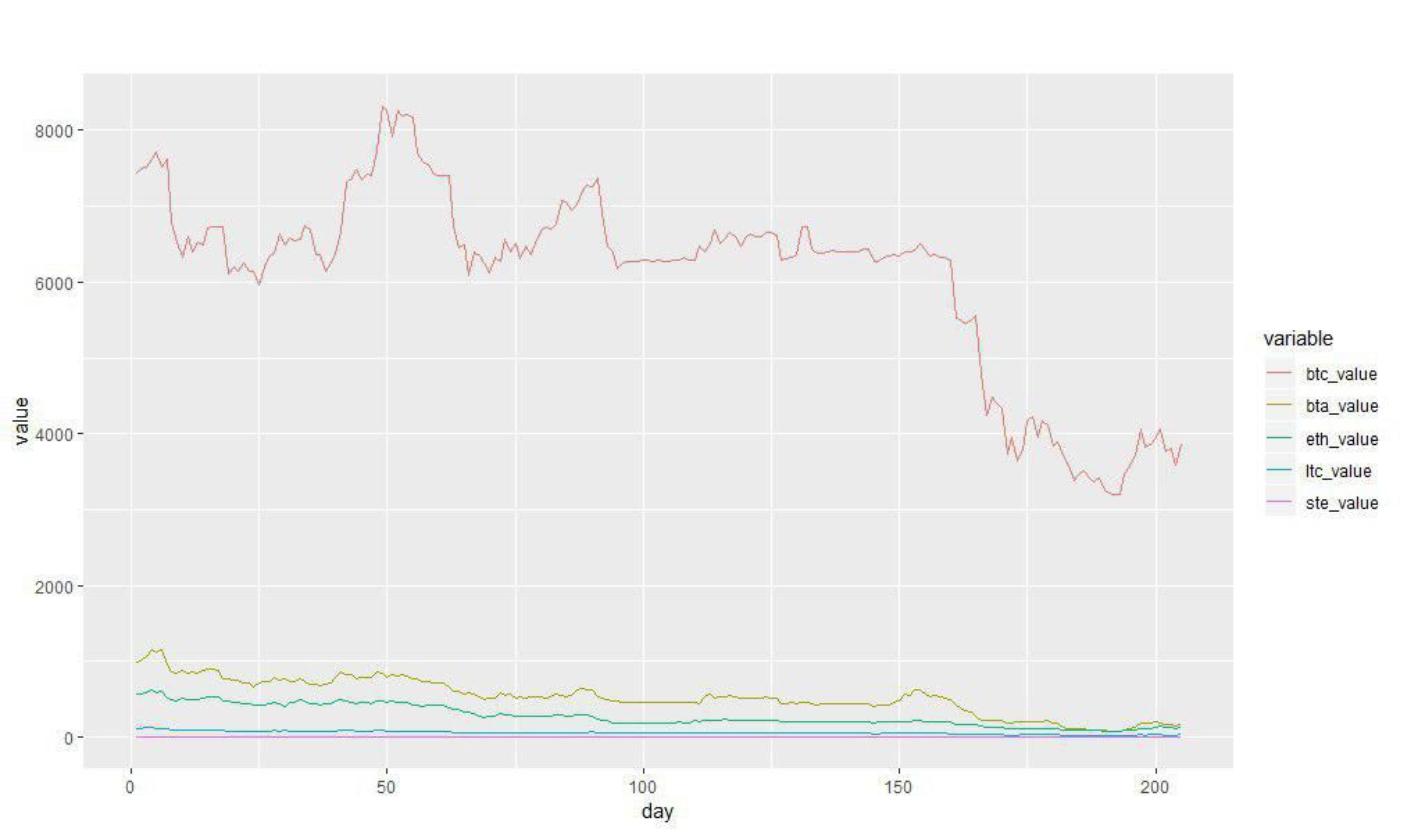} %\\
 \caption{Value in USD of Bitcoin (btc\_value), Bitcoin Cash (bta\_value), Ethereum (eth\_value), Litecoin (ltc\_value), and Stella (ste\_value) in 2018. \label{fig:ValueUSDBitcoinCashEthereumLitecoinStella}} 
\end{figure}

On the other hand, although this research focuses on BTC, because it is the most widely known cryptocurrency, the irruption of different digital assets in the last years has led to a variety of distinct currencies in this new market, with different structures and situations. We reviewed the behavior of the 5 most popular cryptocurrencies: (1) Ethereum, which inherits most of the BTC characteristics, but it is programmable so that it can be used in more applications than Bitcoin; (2) BitcoinCash represents a split in BTC by  increasing the block size limit, so that this cryptocurrency can increase the number of transactions per second; (3) Stellar  which is characterized by the transactions with real currencies, so that it is possible to send a specific amount in one currency and obtain the equivalent in another one; and, (4) Litecoin, the only cryptocurrency that can be complementary to BTC,  is supported by an open-source system that allows for the user to administrate their finances (Litecoin has four times the number of units BTC has). These four different digital assets have different features and they illustrate the variety of cryptocurrencies in the market. Figure~\ref{fig:ValueUSDBitcoinCashEthereumLitecoinStella} compares the different values of these cryptocurrencies when there are similar situations that allow such comparisons. The number of days has been reduced, because some cryptocurrencies do not have data during some months in 2018. There are certain similarities between the first four cryptocurrencies, but the value of Stellar is too small to draw any conclusions.\vspace{-12pt}

A correlation between the five cryptocurrencies and the number of posts and topics reveals some interesting results that require a detailed description. Table~\ref{tab:ClassificationCorrelationClosingValueBitcointalk} shows a surprising improvement in the correlation between the data in the forum and Bitcoin Cash, Ethereum, and Litecoin, where the data are more condensed in a direct relationship. The~activity in the forum includes information that is related to other cryptocurrencies. Besides, the other cryptocurrencies have a static value and their activity is quite similar due to the activity in the Bitcoin market and its value. On the other hand, Stellar is not very known, its value is quite low when compared to the other cryptocurrencies, and its market is not fully developed. Ethereum presents the best results in terms of the reaction from the forum. This~proves that Ethereum is consolidated as an alternative to BTC and the community seems to react positively  to its performance.

\begin{table}[H]
%\tablesize{\small}
\caption{Classification of correlation between closing value and data from \emph{Bitcointalk}. \label{tab:ClassificationCorrelationClosingValueBitcointalk}}
%\setlength{\cellWidtha}{\columnwidth/3-2\tabcolsep+0.0in}
%\setlength{\cellWidthb}{\columnwidth/3-2\tabcolsep+0.0in}
%\setlength{\cellWidthc}{\columnwidth/3-2\tabcolsep+0.0in}
%\scalebox{1}[1]{\begin{tabularx}{\columnwidth}
%{>{\PreserveBackslash\centering}m{\cellWidtha}
%>{\PreserveBackslash\centering}m{\cellWidthb}
%>{\PreserveBackslash\centering}m{\cellWidthc}}
%\toprule
\begin{tabular}{lll}
\textbf{Cryptocurrency}	& \textbf{Correlation Posts}	& \textbf{Correlation Topics} \\ 
Bitcoin	& 0.738818	& 0.7339391 \\
Bitcoin Cash	& 0.8854177	& 0.8888595 \\
Ethereum	& 0.8979489	& 0.9018563 \\
Litecoin	& 0.8831485	& 0.8892847 \\
Stellar	& 0.5354016	& 0.5328425\\

%\bottomrule
\end{tabular}
\end{table}

On another hand, the findings of this study have to be seen in light of some limitations, as follows. From the perspective of unexperienced investors, and as a consequence of the lack of regulation, the most serious risk in cryptocurrency markets is volatility. The big fluctuations in exchange rates directly affect the holders when exchanging the virtual currency into real one. It is also notable in user purchasing capacity, due to price instability. Therefore, cryptocurrencies have been traditionally seen as a speculation vehicle instead of a form of payment, and their use should be discouraged to those who have no or little knowledge about them.  However, the historical idiosyncrasy of the period of our study where cryptocurrencies were only in its infancy should be taken into account. According to a recent study in~\cite{doi:10.1063/1.5036517}, cryptocurrency markets (BTC in particular) are showing similar characteristics to the ones in regulated markets, so a new market is foreseen in parallel to the traditional regulated markets. This observation was confirmed in an ulterior study in~\cite{doi:10.3390/fi11070154} where ETH is also considered. However, in~\cite{10.3390/e22091043}, research observed that, despite the fact that most of the time cryptocurrency market does away with traditional markets, both coupled temporarily during 2020. Although our study focuses on the behavioral and social characteristics of cryptocurrency market, an extensive and up-to-date analysis is needed in order to confirm this trend.  If traditional and cryptocurrency markets are going alike from a financial point of view, the use of social media as a sign of evolution and volatility of cryptocurrencies should be revisited to compare the correlation level between prices and social activity in both regulated and non-regulated/auto-regulated markets. 

Finally, as mentioned, the study shows that forum activity is influenced by cryptocurrency applications. If cryptocurrency forums are used as supporting mechanism for trading in non-financial sectors, the financial activity should be isolated from the potential collateral effects of the usage of cryptocurrencies in different sectors. During the period of our study, this fact was specifically observed in cyber-video games and sports gambling. It~deserves further research to uncover when these activities should be considered in isolation or, on the contrary, the usage of cryptocurrencies in games and gambling is affecting to the trading dynamics in a relevant way, and they should be considered all together. In~fact, in~\cite{Ethereum-Crypto-Games}, the authors analyze the trading of game items for cryptocurrency (ETH) in popular crypto-games and state that the introduction of cryptocurrencies in video-games is a sign of the convergence of digital gaming and gambling. To our knowledge, there is little research in the literature on the implications of cryptocurrency-based gaming and gambling in the financial dynamics of the cryptocurrency market.

\section{Conclusions}
\label{sec:conclusions}

This research focuses on the study of the interplay between the cryptocurrency community Bitcointalk and the fluctuations and trends in the values of BTC. The study states that forum activity  keeps a direct relationship with BTC value and that forum content can explain specific events related with the cryptocurrency market. Specifically, the results show the relevance of forum quotes, since they usually correspond to conversations where users try to sell BTC in scenarios of decreasing value, to weekly concentrations of posts around specific topics or even to discussions about peaks in the BTC value. In this respect, cryptocurrencies can be considered as any other product, so that content that users share on forums, in the form of opinions and observations based on intuitions, can have and impact on product value. However, with cryptocurrencies being a financial product, that impact leads to value fluctuation and insecurity: investors have no guarantees to finish a sale or to  mine massive amounts of Bitcoins, but, at the same time, the cryptocurrency community can put pressure on the market and influence the upcoming values. Further conclusions focus on the users and their participation in the forum. Most of the users in 2018 are new to cryptocurrencies, while a few of them were active since 2013 and they have earned certain prestige thanks to their transactions with Bitcoins. This suggests that dataset segmentation by user types potentially results in a higher correlation, because of the strongest position of prestigious users in determining a peak or a fall in the market value. Finally, the study also shows the role of cryptocurrency forums as support to orchestrated group activity, like the Pump-and-Dump scenario.

Taking into account the potential limitations in this study, as ongoing work, we are  extending the predictive part of the paper by considering recent study periods. This ongoing work would confirm a higher maturity in the cryptocurrency financial forums which would correspond with the evolution of cryptocurrency markets into a more stable scenario. The extended period will also provide the opportunity to test out-of-sample to reach sounder insights and to move froward from cross-correlation analysis to a prediction model. If cross-correlation results indicate that past values of an input series (forum activity) influence the future values of a response series (cryptocurrency price), the next step is constructing a prediction model---based on ARIMA (Autoregressive Integrated Moving Average)  or VAR (Vector Autoregressive) . Also, and according to the recent observation about the behavioral alignment between traditional markets and cryptocurrency market, and taking into consideration the general belief that, at least in traditional markets, investors’ sentiment plays an important role in trading, not only should the activity dynamics be used as a cyber-social predictor for cryptocurrency returns, but also content in the contributions of the financial forum should be researched in order to analyze its predictive power. In~his respect, further work should use the polarity and/or the sentiment of the forum contributions as predictor. Also, these parameters should be combined with credibility and provenance scores. As mentioned before, this work is located into an ecosystem to provide a simple dashboard to small investors. Given the risks in non-regulated markets, mainly cryptocurrencies, provenance and credibility scores should be defined.  In our previous work for regulated market, provenance and credibility can be estimated by analyzing corporative data sources. For the case of cryptocurrencies, aspects like the social reputation of the contributor and also the content and sentiment of the contribution will be taken into account to define provenance and credibility scores which can mitigate investors’ misleading in the cryptocurrency market.

Another research line in our future work focuses on the usage of social  media data sources as part of forensic analysis of criminal activities, which should generate insights for a better safeguard of the financial system in the presence of virtual currencies. On the other hand, cryptocurrency pseudonymity attracts criminals to carry out their illicit activities i.e., illicit goods and services on Darknet markets, ransomware attacks, extortion and money laundering. Moreover, peer-to-peer authentication in cryptocurrency transactions, that bypass institutional intermediaries, dismantle the global AML (Anti Money Laundering) schemes in traditional financial markets. As in this paper, behavioral science can be applied to social media  data sources to uncover criminal activities.  Content in cryptocurrency-specific forums or general-purpose financial forums is considered relevant as evidence of money laundering and so can be used as input to flexible AML schemes,  but it should be accompanied by the analysis of cryptocurrency payment flows to support a forensic method for illicit activities.

\end{document}